\documentclass[12pt,a4]{article}
\pdfoutput=1

\usepackage{graphicx}

\def\thesection{\arabic{section}.}

\def\appendix{\setcounter{section}{0}
        \def\thesection{Appendix.}
        \def\theequation{\Alph{section}.\arabic{equation}}}
\def\IR{{\hbox{{\rm I}\kern-.2em\hbox{\rm R}}}}
\def\IH{{\hbox{{\rm I}\kern-.2em\hbox{\rm H}}}}
\def\IC{{\ \hbox{{\rm I}\kern-.6em\hbox{\bf C}}}}
\def\IZ{{\hbox{{\rm Z}\kern-.4em\hbox{\rm Z}}}}

\newcommand{\beq}{\begin{equation}}
\newcommand{\be}{\begin{equation}}
\newcommand{\eeq}{\end{equation}}
\newcommand{\ee}{\end{equation}}
\newcommand{\bea}{\begin{eqnarray}}
\newcommand{\eea}{\end{eqnarray}}
\newcommand{\bean}{\begin{eqnarray*}}
\newcommand{\eean}{\end{eqnarray*}}
\newcommand{\ba}{\beq\begin{array}{lll} }
\newcommand{\ea}{\end{array}\eeq}
\newcommand{\bi}{\bibitem}

\def\IC{ {\rm l\hspace{-1.2ex}C} }    
\def\IZ{{\hbox{{\rm Z}\kern-.4em\hbox{\rm Z}}}}
\def\IR{{\hbox{{\rm I}\kern-.2em\hbox{\rm R}}}}
\begin{document}                           %
                                                                 %
                                                                 %
%
%
%
%
\def\citen#1{%
\edef\@tempa{\@ignspaftercomma,#1, \@end, }
\edef\@tempa{\expandafter\@ignendcommas\@tempa\@end}%
\if@filesw \immediate \write \@auxout {\string \citation {\@tempa}}\fi
\@tempcntb\m@ne \let\@h@ld\relax \let\@citea\@empty
\@for \@citeb:=\@tempa\do {\@cmpresscites}%
\@h@ld}
%
\def\@ignspaftercomma#1, {\ifx\@end#1\@empty\else
   #1,\expandafter\@ignspaftercomma\fi}
\def\@ignendcommas,#1,\@end{#1}
%
%
\def\@cmpresscites{%
 \expandafter\let \expandafter\@B@citeB \csname b@\@citeb \endcsname
 \ifx\@B@citeB\relax 
    \@h@ld\@citea\@tempcntb\m@ne{\bf ?}%
    \@warning {Citation `\@citeb ' on page \thepage \space undefined}%
 \else
    \@tempcnta\@tempcntb \advance\@tempcnta\@ne
    \setbox\z@\hbox\bgroup 
    \ifnum\z@<0\@B@citeB \relax
       \egroup \@tempcntb\@B@citeB \relax
       \else \egroup \@tempcntb\m@ne \fi
    \ifnum\@tempcnta=\@tempcntb 
       \ifx\@h@ld\relax 
          \edef \@h@ld{\@citea\@B@citeB}%
       \else 
          \edef\@h@ld{\hbox{--}\penalty\@highpenalty \@B@citeB}%
       \fi
    \else   
       \@h@ld \@citea \@B@citeB \let\@h@ld\relax
 \fi\fi%
 \let\@citea\@citepunct
}
%
%
%
%
%
%
\begin{titlepage}
\vspace{.5in}
\vspace{3.5in}

\begin{center}
\ 
\vspace{1.0in}

\ 

{\Huge\bf Information, complexity, brains and  reality \\   }

{\huge\bf (Kolmogorov Manifesto) \\[1.2in]   }

\includegraphics[width=4cm]{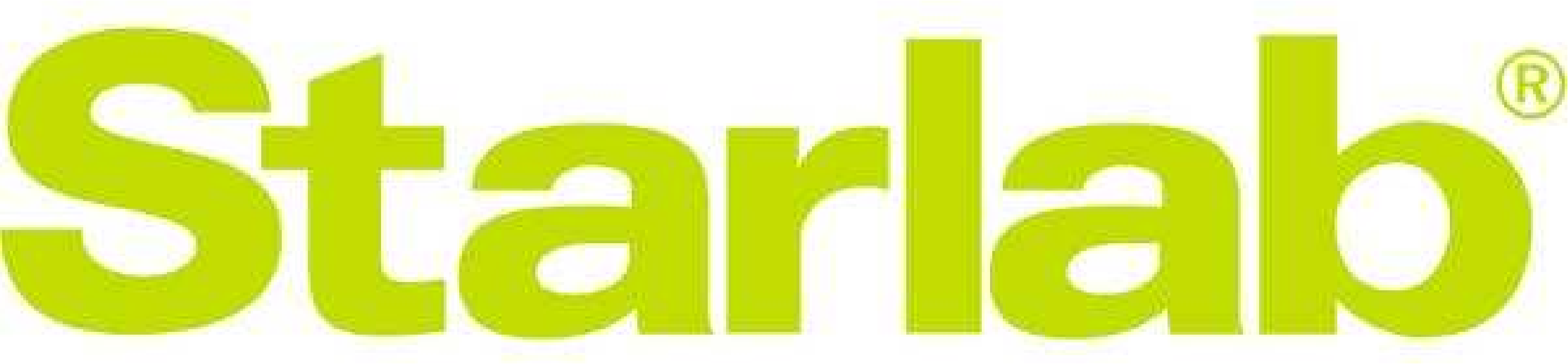}\\ \vspace{.4in}
{\bf Starlab Technical Note TN00054} \\
{\bf Status: Public}\\
{\bf Initiated: 06-2001}\\
{\small  Revisions: 10-2002, 12-2002, 12-2003, 09-2004, 12-2004, 02-2005, 08-2005, 10-2005, 07-2006, 12-2006, 04-2007}\\
\vspace{.4in} 
   {Giulio ~R{uffini}\footnote{\it email: giulio.ruffini@starlab.es}\\
        {\small\it Starlab}\\
        {\small\it Edifici de l'Observatori Fabra, C. de l'Observatori s.n.}\\
{ \small\it Muntanya del Tibidabo, 08035 Barcelona, Spain \\ }
        } 
 \end{center}

 \vspace{.2in}

\end{titlepage}
\   

\clearpage 

\tableofcontents

\clearpage

\  \clearpage 

\section*{ABSTRACT}
 In these notes  I discuss some aspects of information theory and its relationship to physics and neuroscience. Information has a rather central role in all human activities, including science, and it is a well defined concept associated to the number of degrees of freedom of a state. The unifying thread of this somewhat chaotic essay is the concept of {\bf Kolmogorov or algorithmic complexity} (Kolmogorov Complexity, for short). Another important notion in the discussion is that of a  {\bf Self-Entity or Agent}, which  is defined here as a semi-isolated physical (or informational) system capable of controlling its physical interfaces or couplings with the rest of the universe. This working definition may be thought of as the conceptual primitive for the brain.  
\medskip

I analyze implications of  the assertion that the brain is a modeling tool exchanging information with the rest of the  {\bf Universe} and conclude  that,  ultimately, all questions about reality should be framed in an information theoretic framework with the brain at the center. Indeed, by ``Universe'' here I  mean the universe of sensory input to our brains,  the only acceptable starting point for the definition of {\bf Reality}, and that for all practically purposes one can state that there is only {\bf information}. As a consequence,  any theory of everything must start from information theory. Put in other words, information theory provides the conceptual framework to understand the relation of brains with the Universe. In particular, one of the building blocks of physics is the concept of state. Information is the currency of the brain to talk about states. 
\medskip

As a consequence of this view, I will try to show that it is natural to interpret cognition, in general, and science, in particular, as the art of finding algorithms that apprach the Solomonoff-Kolmogorov-Chaitin (algorithmic) Complexity limit with appropriate tradeoffs. In addition, I argue  that what we call  the universe is an interpreted abstraction---a mental construct---based on the observed coherence between  multiple sensory input streams and our own interactions (output streams). Hence,  the notion of Universe is itself a model. {\bf Rules} or regularities are the building blocks of what we call Reality---an emerging concept. I highlight that physical concepts such as ``mass'', ``spacetime'' or mathematical notions such as ``set'' and  ``number'' are models (rules) to simplify the sensory information stream, typically in the form of invariants. The notion of {\bf repetition} is in this context a fundamental modelling building block. Although the discussion is quite high level and verging on philosophy, I try to illustrate these notions with examples from EEG experimental data analysis with subjects realizing low level pattern learning tasks, where repetition plays a central role. 

\medskip

As it turns out, these ideas can be framed naturally in the context of an emerging field called {\bf Presence}. Presence studies how the human brain constructs the model of reality and self using replacement/augmentation of sensorial data. Presence 
belongs to a wider class studying how cognitive systems build models of their environment and interact with it. In Presence, the technological goal is to place a brain in a controlled and convincing and interactive information bath.
\medskip

In the later sections I consider the fact that our brains have access to {\bf incomplete information} about the state of the universe, and discuss its implications in the theories of statistical and quantum mechanics, both of which have this much in common: they deal with incomplete knowledge about the classical microstate in some sense. I emphasize that the so-called macrostate is not a state at all.  
\medskip 

In relation to biology and the origins of the brain and in the same vein, I argue that life and, in particular, nervous systems evolved to find methods to simplify and encapsulate the apparent complexity of the universe, the context  being the usual one: permanence of the fittest (natural selection). Thus, biological evolution is Nature's algorithm for  the optimal compression problem, a fact that we can seek to exploit both in research and in practical applications.  \medskip

Finally, I try to address the problem of time in the same light. I discuss briefly the role of time in computation and evolution, and consider a timeless, computer-less description of reality in which all that exists is a program.
\clearpage

\section{INFORMATION}
\setcounter{footnote}{0}

In these leisurely written notes, I analyze and discuss the implications of a realization I had while thinking about a particular experiment in a well-known and fastinating neuroscience experimental paradigm. In this experiment\footnote{Experiment  by the team of Neurodynamics at the University of Barcelona (UB-NL) led by  Dr. Carles Grau.}, further discussed below, subjects were presented so-called ``standard'' and ``deviant'' auditory patterns (i.e., controlled sound sequences delivered through headphones), and their electroencephalograpic  responses recorded on key with the presentation of the patterns. This is  a standard experimental set up to study the response of the brain and nervous processing to such stimuli using electroencephalography (EEG). EEG is a fascinating measurement technique exploiting the fact that our nervous system relies on  electrical signals to communicate and process information. EEG is a tool providing fast temporal sampling of brain activity (as opposed to others such as functional magnetic resonance imaging).  The experiment, which I describe below, and related discussions led  me to think about the brain as a pattern discovery machine seeking to model and therefore compress the input information stream. What we call modeling is in fact directly related to the concept of algorithmic complexity---a concept I had not explicitly encountered at the time but which intuitively I was looking for. I came across a book which discussed the issue of algorithmic complexity and, more importantly, defined Kolmogorov complexity, a very beautiful concept which is the unifying thread in the following sections \cite{Cover}. 	 		

\medskip

Putting the brain in center stage together with the idea that brains seek to model and simplify the incoming flux of information has important consequences. I will try in these notes to place this perspective in some sort of a theoretical context. The motivation is the deep connection between physics, information science and cognitive systems (of which brains are a prime example).

\begin{figure}[t!]
\label{brain}
\includegraphics[width=15cm]{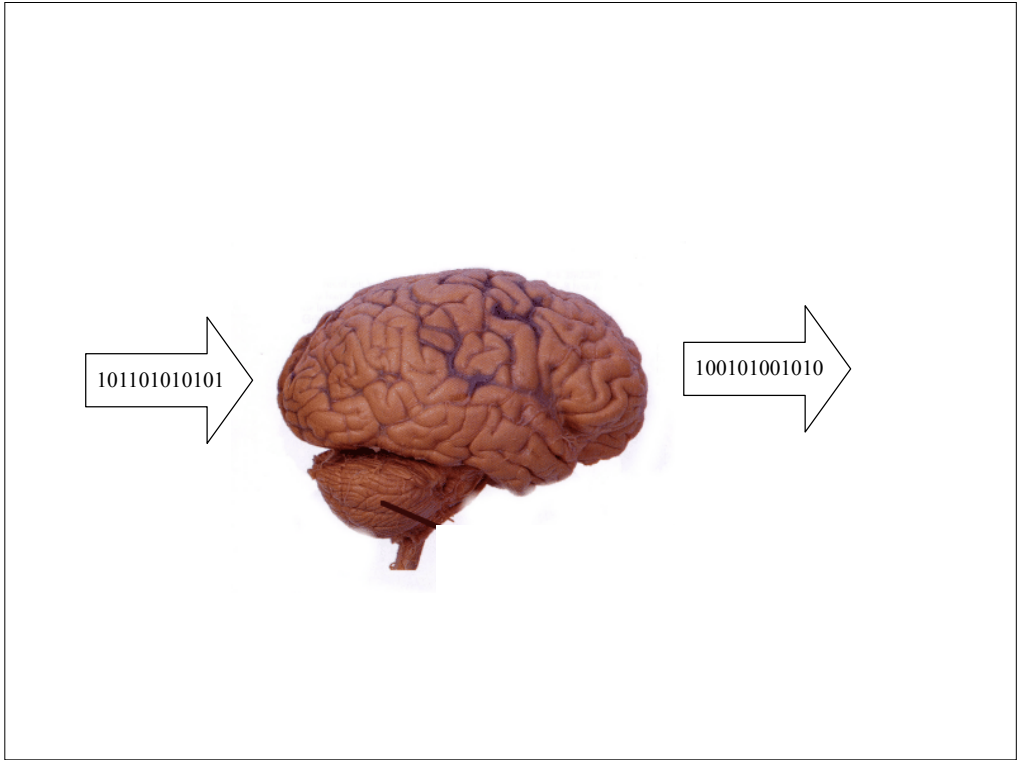}
\caption{The central concept in this paper is this: the brain creates the model of reality (including self) through  information interaction (in and out) with the universe ``outside''. The universe outside includes any intermediary body. The universe outside is the ``information bath'' }
\end{figure}
\medskip

For example, consider the following question: what is Science? Surely, we all have an idea of that. Science is what scientists do...Physics, Biology are clear examples of scientific disciplines. Science is about observing phenomena and understanding nature, about  extracting the rules of reality in  different disguises. Science can and should be used to improve our lives. These are all terms of reference for science. Yet, we can give a more concise definition of science. A simple and intriguing answer is that science is what brains do. This should be taken as a definition of science---not a definition of what brains do, although in a sense they are equivalent.  What this says, in essence, is that we are all scientists, whether we know it or not.    As I discuss below,  brains  try to simplify the incoming information flow gathered by our senses. Simplicity is power---this much should be rather clear. And I plan to walk further down this conceptual road and  convince the reader that in some sense ``simplicity'' is actually ``reality''.

\medskip 
Science, in the traditional sense of the word, is of course also about seeking simplicity and compression. When a scientist  develops a model of what is observed, the result is a tool for data compression. A good model can be used to literally compress the observed data. Just compute the difference between data and model and compress the result. A good model will give result in a difference file with mostly zeroes.   \medskip

Although it might be true that we all are scientists, science requires further discipline and a rather strict methodology. It could be called ``professional thinking''. But in essence, the difference between a Scientist and a scientist is quantitative, not qualitative.

\medskip

What does this all have to do with reality? What is the essence of reality? This is a deep and very old question, asked for the first time a long time ago. If you are reading this, you definitely exist. What is reality for you? I would like to argue that the concept of  reality can be firmly anchored in the concept of information---and that our perception of reality is the model we build to simplify our existence based on that information. {\bf Reality is a construct of our brains---a model}. Reality is the  model brains build to sustain existence. 

\medskip

How does the brain create this construct? If we don't understand this it will be very hard to progress in fundamental science. We have already encountered stumbling blocks that elude us, such as the role of the observer in Quantum Mechanics. I think these issues are profoundly related.
\medskip

Here are some somewhat provocative statements related to the discussion and to be  further ananalyzed below:
\begin{enumerate}
\item Reality is a byproduct of information processing. Reality is a model our brains, swimming in an information bath, build to sustain existence.
\item Mathematics is the only really fundamental discipline and its currency is information.
\item The nervous system  is an information processing tool. Therefore, Information Science is crucial to understand how brains work.
\item The brain compression efforts provide the substrate for reality (the brain is the ``reality machine'').
\item The brain is a pattern lock loop machine. It discovers and locks into useful regularities in the incoming information flux, and based on those it constructs a model for reality.
\item Definition: A {\bf self-entity or agent} is a physical semi-isolated system  capable of selecting and partially controlling physical interfaces (``couplings'') with the environment. It is the physical version of a Turing Machine. 
\item Definition (alternative version): A {\bf self-entity or agent} is an informational semi-isolated system  capable of selecting and partially controlling information interfaces (``couplings'') with the environment. 
\item A self-entity has only access to {\bf partial} information about the universe.
\item Information flows between self-entities only through (physical or informational) interactions which we call sensors and actuators (interfaces).
\item The mutual information between coupled systems is nonzero (this is a consequence of physical law, e.g., quantum mechanics).
\item Pleasure and Pain (to be further defined) are the sole drivers of action, the goal functions. Our ``firmware''.
\item Genes contain (mutual)-information about their targeted habitats.
\item A successful genome will have large mutual information with respect to its environment.
\item The information flux into the human brain is of the order of\footnote{This is just a wild guess: I want to bring the questions to life, but the answer is far from obvious.} 1 GB/s
\end{enumerate}
The first two statements, for example, provide a tentative answer to the problem of the {\em unreasonable effectiveness of Mathematics in Physics}: Mathematics is effective at describing reality simply because reality is information, and mathematics is basically dealing with the management of  information and modelling.
\medskip

In what follows we will suppose that we are physical instantiations of Turing machines, that all we really know about the universe comes through our {\bf senses} and interaction via {\bf actuators} in the form of information, and that brains seek to compress this information in a useful manner---and see where this leads us. \medskip

Another aspect I will bring to the discussion is the problem of Time. What is Time? Why do we "feel" time the way we do? For example, why do we have memories of the past, but no memories of the future if physical laws are time-reversal invariant? Where does the arrow of time come from? Information and computation seem to require both dynamics and therefore,  Time. Computation and time flow go hand in hand, and it is hard to picture one without the other. Can we learn something by looking at this old problem from an information theoretic point of view?
\medskip

Finally, let me clarify that the ideas  in this paper are mostly original (to the extent that anything can be original), although  by now I am fully aware that other people are working along similar lines and I am sure thinking very similar thoughts. Since I started working on this notes, books like ``On Intelligence'' \cite{hawkins2004} and more recently  ``The Computational Universe'' \cite{lloyd2006} have come out, both of which deal with the issue of algorithmic complexity, and even a movie (``What the Bleep'').  This is probably a consequence of the fact that we have now entered in full the Information Era, where information and brains have come into sharp focus. As I mentioned above, I encountered the notion of Kolmogorov Complexity  when looking for a mathematical definition to describe  the capacity of the brain to detect and learn patterns. I was amazed and excited by how well this powerful concept fit the notion I was looking for. I am sure these ideas will resonate strongly in other minds.  \medskip 

Finally, I would like to clarify that the nature of this work is rather exploratory, a personal tour of some of the implications of information science and algorithmic complexity. I would have never started writing it if I'd waited for the moment in which it was all clear. Alas, this moment may never come!  For all practical pursposes, then, this is a live document and I intend to continue working on it and revising it. 

\medskip 
Hopefully, in the meanwhile the  reader will find something useful in it and take it along.

\subsection{Information and Sensors}
\label{sensors}
Information is that which  allows us to fix, completely or partially, the ``state" of a system. By ``state'' and ``system'' we refer here to the standard physical concepts. 
Information theory was born many centuries ago (think of cryptography) but came of age with the advent of the telegraph and electromagnetic communications. I think that physicists have been using this concept for centuries, because the notion of "state" necessitates a conceptual framework using information. In physical terms, information is what is needed to constrain and ultimately fix a microstate, and it is easy to see that one can do that in a succint or redundant manner.
\medskip

Let me give an example of what concretely a state is and how you would go about describing it with information. Consider a 2D universe with 3 undistinguishable particles. Then, the classical state of the system could be described by the following expression:
$$
S=\{[p_x^1, p_y^1,  v_x^1, v_y^1] , \, [p_x^2, p_y^2,  v_x^2, v_y^2],\,  [p_x^3, p_y^3,  v_x^3, v_y^3]\}.
$$
You can read this as saying, ``there is a particle at position such and such and velocity such and such, another one at...'' etc. Each one of the entries is a number, and you have to decide what precision you want in each. Classically, we can just say that we need to provide 9 degrees of freedom. The classical concept of degree of freedom was closely related to that o information. In order to fix the state, a classical observer is allowed to make as many measurements as she likes. Measurements will disturb a bit the state, but in principle the error can be made as small as desired. So a classical observer in a classical world can get all the information describing the state of the observed system. \medskip

In quantum mechanics the universe is also described by a state, and in this case the state is a function---the wavefunction. Each pair of classical {\bf degrees of freedom} becomes a variable in this function. So in case above, you would have to specify a state of the form 
$$
\Psi=\Psi(x_1, x_2, x_3).
$$
Furthermore, this function is symmetric in the three variables if the particles are undistinguishable. In this case, however, measurements will disturb the state. The observer can access some aspects of the state (technically, she has to choose some basis), but not all. A quantum measurement provides information about a component of the state in some chosen basis, and destroys the rest. Quantum mechanics rules are really about what information an observer can access on the universe.

\medskip

In this light, one could say that  classical and quantum mechanics theories  regulate how an agent can capture information from the universe or observed system, and that they do so in a very different manner. But information remains at the center stage in both. In fact, I would say that the limitations imposed by quantum mechanics highlight even more the role of information in physical  theories. And perhaps more significantly, the central role of the observer in the story.

\medskip 

Imagine an agent (human, machine) which is inmersed into an environment. The agent has programs which allow it to cope with the environment and thus survive. In order to put them to work, though, it needs {\bf information}. It needs to assess, to some extent, the state of the environment. Through sensors, it recovers information from the environment, including the state of its  own `''interfaces'' (``In what position  are my arms?''). 

\medskip 
In order to discuss states and information exchange, we need to carefully define what  a sensor is. 
For the purposes of this work, I would like to propose the following information theoretic definition of a  {\bf sensor}: 
\begin{quote}
{\em A sensor is a device that can capture and relay information from a sensed system to a commanding system}. 
\end{quote}
Note that according to this definition, certain objects or systems which we may not expect to have such a status will also fit the definition of sensor. For instance, a USB flash memory stick is a sensor, according to this definition, as is a cellphone! In my view, these are perfectly good examples of a sensor. Another interesting way to state the same is that there is an increase in the mutual information between the sensor and the sensed system\footnote{Is this the Heisenberg Principle in disguise? When we measure something we disturb it because the sensed is also ``tainted" by the sensor through this information exchange. This process seem rather fundamental and unavoidable.}. 
When we say that a sensor transduces all we are saying is tha the information is encoded in the state of some physical system first, then in another. But the bottom line is that a sensor is there to capture information.  \medskip

\begin{figure}[t!]
\label{brain2}
\includegraphics[width=15cm]{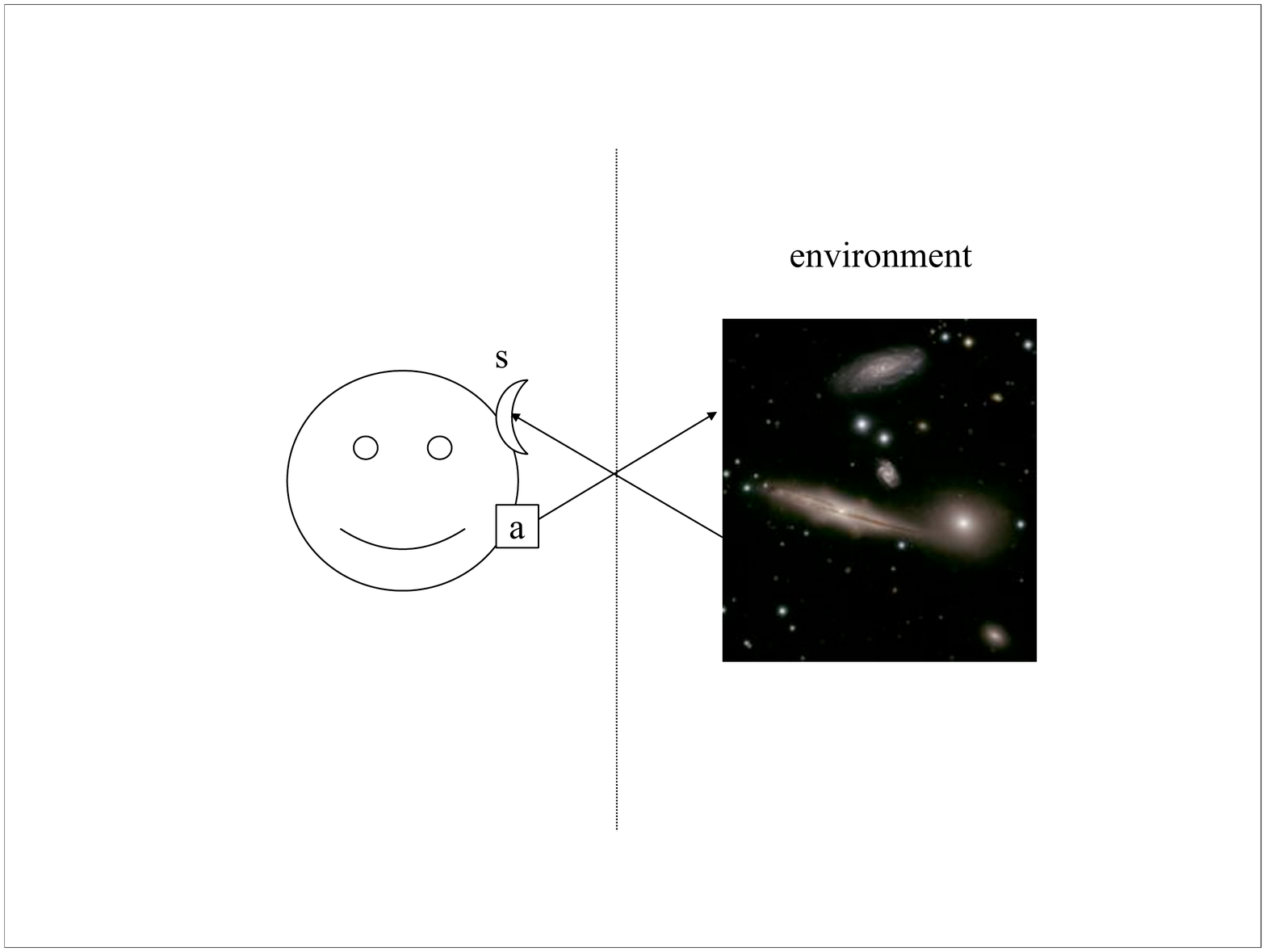}
\caption{The brain creates the model of reality through  information exchange (in and out) with the universe ``outside''. In this case we show a brain interacting with the normal universe by bi-directional information exchange, using {\bf sensors} and {\bf actuators}.}
\end{figure}

Thus, sensors here are just information interfaces between the universe and the entity. This interfaces may be passive or active (sensing can be a rather active process, using computational feedback to focus on different information streams). The universe may be in any of many states. In general, a sensor  provides only  {\em partial} information about what state that may be, especially in quantum mechanics (more on this topic below). The combination of several sensors provides the means for a sharper identification of the state (through data fusion).
\medskip

It is more than conceivable that the evolution of sensors by entities is closely related to the ability for model building. Some sensors seem to work better than others, capturing relevant information for survival. From the evolutionary point of view, if some information streams add little value, the associated sensors must be ultimately discarded. In fact, this can also happen  at the organism level. Sensor selection, sensor focus, is  already part of the compression problem as approached evolutively. Sensor selection is equivalent to a particular choice of lossy compression.  Of course, sensors are necessary but not sufficient.  In order to have an interesting, interacting entity,  sensors, ``computers'', operating systems and programs are needed. Last, but not least, actuators are  highly desirable! They close the loop with the universe, and they allow us to alter it as well as place our sensors in appropriate places (by, e.g., locomotion). One way to look at locomotion (e.g., walking) from the point of view of actuators, is that when we walk, our actuators are sending information to the universe so that it moves and we can access other data streams.  \medskip

Finally, it is worthwhile developing new sensors. By this we mean devices which can capture aand constrain the state in different and more precise ways than our senses. A microscope is an example of this, as is a voltmeter. A more interesting concept is ``sense synthesis''. By this I mean the creation of direct interfaces from the universe to the brain, bypassing our traditional senses and even the peripheral nervous system. Some of this have already been developed and tested, e.g., brain implants into the visual cortex to provide a sense akin to sight\footnote{http://en.wikipedia.org/wiki/Brain-computer\_interface}. However, one can image the creation of totally new senses and perceptions by such direct access to the central nervous system, e.g., to perceive the state of a network, or of a mathematical matrix, or of a community.
\medskip 

\subsection{Kolmogorov Complexity, Science and Evolution}
Here is a naive definition of science: science is what brains do. As scientists, we seek algorithms that lower the apparent complexity of our universe of sensory input. I apologize for the repetition, but I want to emphasize that here by ``Universe''  I just mean  the universe of sensory input to our brains.  The brain is some sort of a box into which a myriad signals arrive from all the sensory input (at what rate?) our body generates after interaction with the universe and from which information is ``transmitted back'' to the universe. 
And I would like to argue that the brain evolved to find methods to simplify and encapsulate the apparent complexity of the universe, the arena being the usual one: survival of the fittest. Simple ``truthful'' models empower agents with better chances for survival and  greater control of our environment. 
\medskip

\begin{figure}[t!]
\label{brain3}
\includegraphics[width=7.5cm]{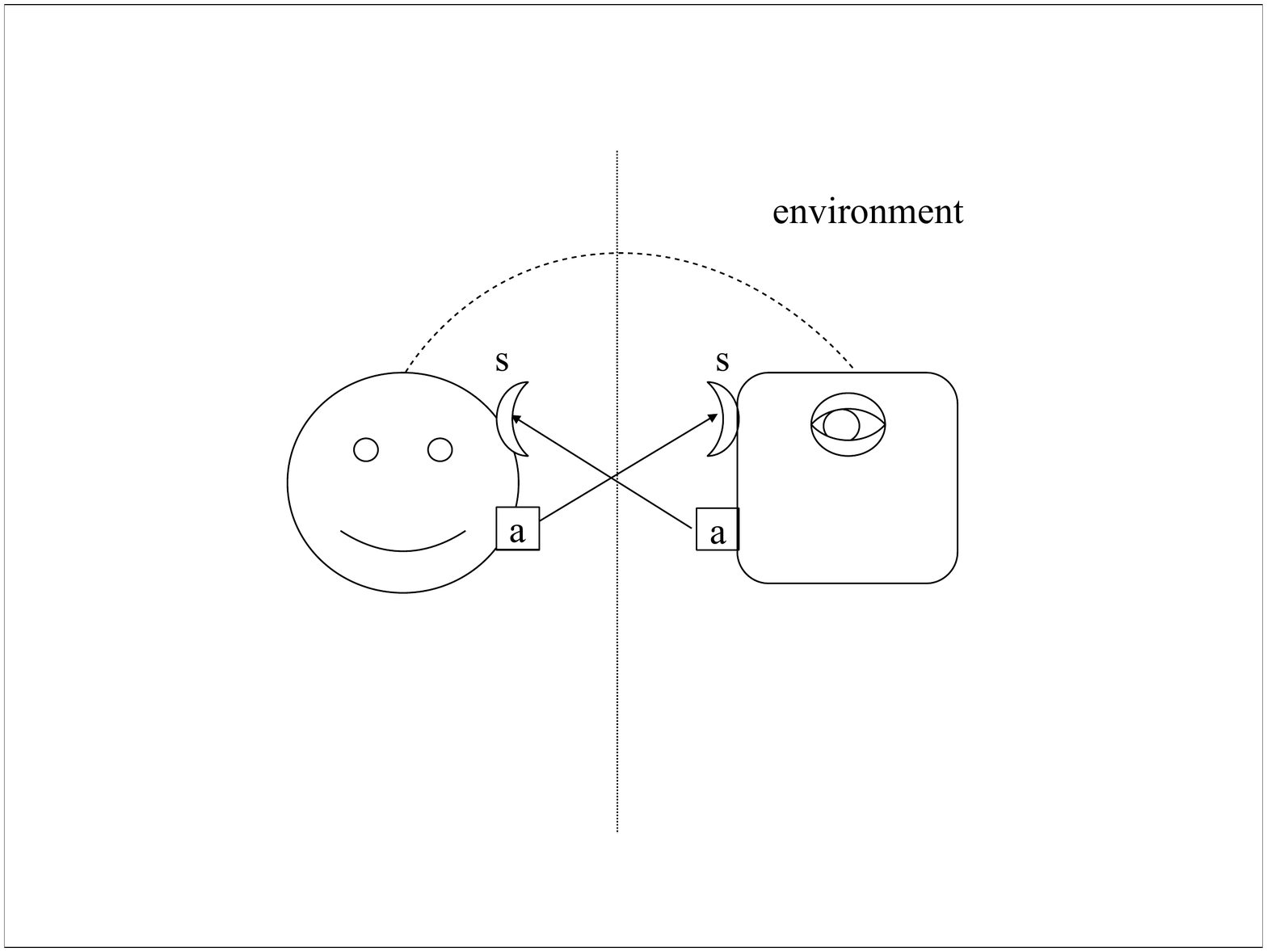} \includegraphics[width=7.5cm]{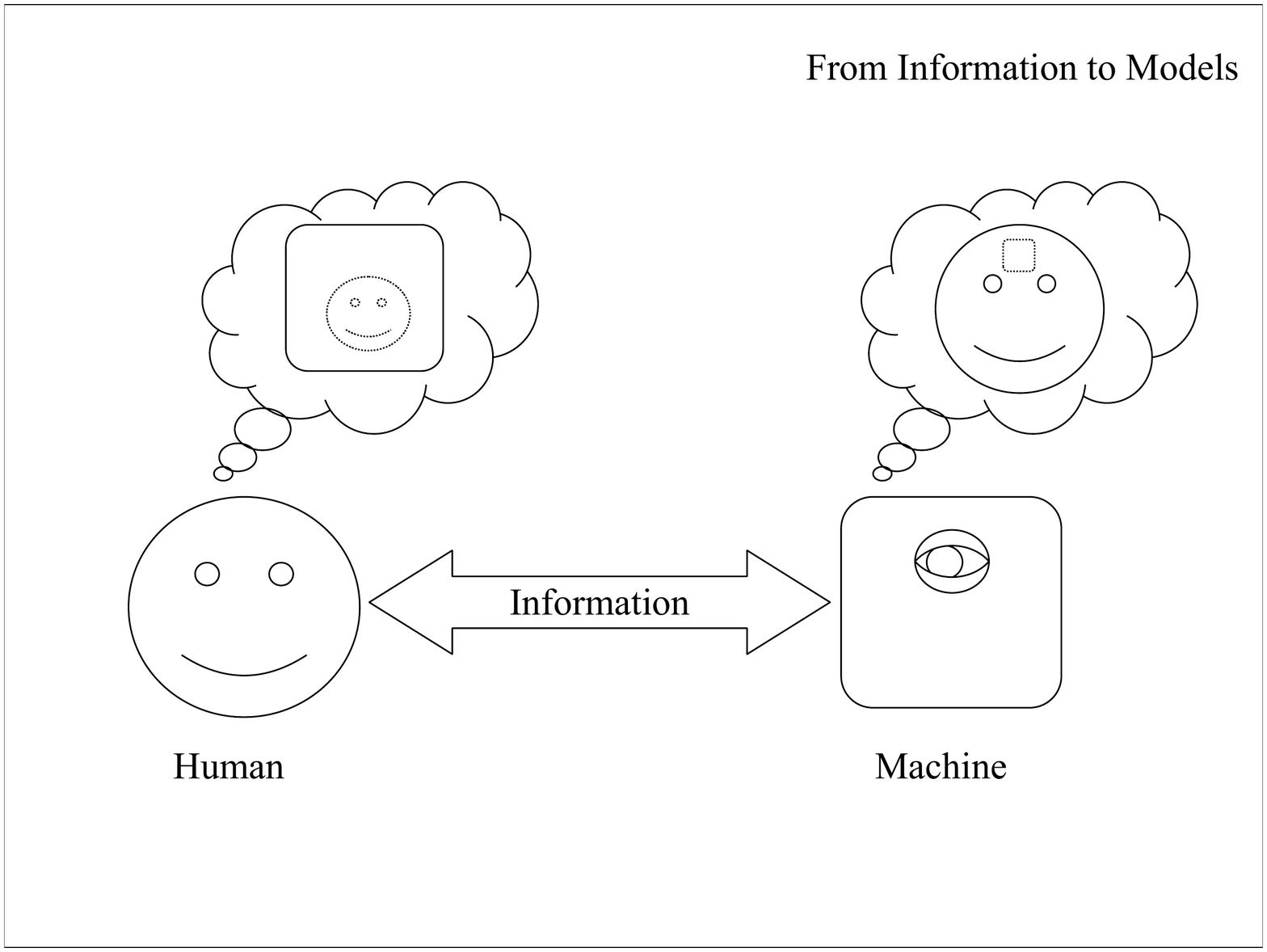}
\caption{The brain creates the model of reality through  information interaction (in and out) with the universe ``outside'' using sensors and actuators. If we ``hack'' the inputs and outputs to the brain with an intelligent, pursposeful information stream, we can put this brain, i.e., fool it into believing,  in an artificial environment. There is also the very interesting possibility of bypassing the senses and going directly from machine to brain in both directions. }
\end{figure}

I order to clarify this somewhat radical information-brain centered position, I would like to consider the following thought experiment. A person is placed into an enviroment in which computers prepare and control her  sensorial input for the person (via sophisticated 3D displays, audio, haptics, smells, vestibular stimuation, etc.) and in which the person brain commands are intercepted at the level of peripheral nerves.  If this experiment is done correctly, the person can be embedded or inmersed into a Virtual Environment. In this experiment, it suffices to describe the universe in terms of bits, because the universe is really a program in a computer managing the sensorial input---see Figure~\ref{brain3}. This experiment has been realized with simple organisms and is the central paradigm in the field of Presence, further discussed below. The reader familiar with the movie Matrix will easily recognize the concept. But I want to emphasize here the central role played by information.  In fact we can recognize three main aspects:
\begin{itemize}
\item a human agent
\item bydirectional human-machine interfaces
\item a machine agent
\end{itemize}
All these have to be there in order to create the subject-universe loop.
\medskip

We recall here  the definition of Kolmogorov Complexity. {\bf The Kolmogorov complexity of a data set is the length of the shortest program which is able to generate it.} Solomonoff, Kolmogorov and Chaitin showed during the second half of the 20th century that this is a good, meaningful definition. In particular,  although the precise length of the minimizing program depends on the programming language used, it does so only up to a constant. We must keep an eye on this aspect also: some programming languages may be more efficient than others, and easier to implement! How is the programming  language itself selected? Here, aspects other than simplicity will no doubt play a role, such as the practical issues of implementation in ``wetware''.
\medskip

As a first example of this concept, consider the sequence
\begin{verbatim}
       1212121212121212121212121212121212121212
       1212121212121212121212121212121212121212
\end{verbatim}
 It is easy to see that its Kolmogorov Complexity is rather small. Here is a simple algorithm to describe it:
\begin{quote}
       ``{\tt Repeat 12 forty times}''
\end{quote}
The fundamental property of the definition of Kolmogorov Complexity is that if we were to write this algorithm in another programming language, it would only differ in length by a constant. By constant, we specifically mean that this number will not depend on the algorithm we wish to translate. In effect, this constant is the length of the translating program (the one translating one language into another). Intuitively, we expect that the above sequence can be coded in a similar manner in different languages.
\medskip 
 
As was mentioned in the introduction and the above example illustrates, the concept of ``repetition'' is a fundamental block in ``compression science''. This concept is rather important and leads brains naturally to the notion of number and counting (and probably mathematics itself). Even simple programs like a  file compressor (e.g., {\it gzip}), will detect this repetitive aspect in the data and compact a large file consisting of repeated 12s rather efficiently. Some very basic blocks of mathematics can be seen arising from the branch of information theory that deals with repetitive phenomena. Come to think of it, the notion of repetition leads to set theory, and counting leads naturally to the concept of number, primes, etc. 
Moreover, repetition is the building block of Recursive Functions, which are indeed known to be equivalent to Turing machines.

\medskip

As a familiar illustration of the fundamental nature of repetition consider this definition: a number is said to be prime if it cannot be (nontrivially) obtained  by the repeated sum of other numbers. That is, if it cannot be expressed as the repeated sum of any other number other than 1 and itself (trivially once). 
That is, it is not possible to generate a prime number by repeating another number many times and summing.
This implies that prime numbers  generate  a clone of $Z$ embedded in $Z$ (the structure of the set  of integers) in the same way that 1 generates Z, and in an isormophic way. That is, 
$ Z \sim p\cdot Z$. The prime $p$ plays the role of the ``indivisible'' 1.  \medskip

Consider now the hypothesis: {\em the number of prime numbers is finite}. In other words, there exist a set of numbers, S, such that the entire number system can be generated from them through multiplication. Well, we know this is not true (the number of prime numbers is infinite). This is similar to the fact that any formal logic system is imcomplete: the analogy is provided by the prime numbers as axioms. An infinite number of ``axioms'' is needed to fill $Z$.

\medskip
Compressive processes are probably very basic mechanisms in nervous systems.  As mentioned above, I would also like to argue that counting, which is at the root of mathematics, is just such type of compactification. Simple repetition is perhaps the most common ``pattern'' detected  by the brain. Counting is just the next step  after  noticing repetition. And with the concept of repetition also comes the notion of multiplication and primality. More fundamentally, repetition is the core of algorithmics.

\medskip
A really interesting definition of mathematics which is very fitting in this paper is provided in 
\cite{Nagel2001}: {\em mathematics [...] is simply the discipline par excellence that draws the conclusions logically implied by any given set of axioms or postulates}.  In other words, mathematics provides the necessary tools for model building. I think this is a very nice and direct definition for mathematics, and it clarifies the role of mathematics in the human brain. Since we hold in this paper that the brain is a model builder, a compression machine, mathematics provides the fundamental substrate to do that, because compression is the reduction of a large set of data or conclusions into a small program or set of axioms.

\medskip
 
Not all instances of information compression are so straightforward. Here is a more challenging case: consider the  data sequence
 \begin{verbatim}
     S=1415926535 8979323846 2643383279 5028841971 6939937510
       5820974944 5923078164 0628620899 8628034825 3421170679
       8214808651 3282306647 0938446095 5058223172 5359408128
       4811174502 8410270193 8521105559 6446229489 5493038196
       4428810975 6659334461 2847564823 3786783165 2712019091
       4564856692 3460348610 4543266482 1339360726 0249141273
       7245870066 0631558817 4881520920 9628292540 9171536436
       7892590360 0113305305 4882046652 1384146951 9415116094
       3305727036 5759591953 0921861173 8193261179 3105118548.
\end{verbatim}
Nothing seems to repeat in a simple manner---but it does! The reader  will no doubt  realize this is the numerical sequence for $\pi-3$.  Euler found a nice expression for $\pi$ which we can use to construct an algorithm to reproduce  this sequence (although we will have to implement an algorithm for taking square roots, summing, etc.):
$$
S=\sqrt{6 \sum_{n=1}^\infty {1\over n^2}} -3.
$$ 
To obtain $\pi$ to a given digit of accuracy, we need to compute a finite number of calculations and sums. To go beyond, the process is repeated. Thus, the algorithm to describe the above sequence (or, more importantly, an arbitrary longer version of it) is rather short. Recently, a digit-extraction algorithm for $\pi$ in base 16 has been discovered \cite{bailey},
$$
\pi=\sum_{n=1}^\infty \left({4\over 8n+1}- {2\over 8n+4} - {1\over 8n+5} -{1\over 8n+6} \right)\left({1\over 16}\right)^n .
$$
Using this algorithm (a repetitive process!) the authors have managed to show, so far, that $\pi$ is random in base 2 (i.e., normal\footnote{http://crd.lbl.gov/$\sim$dhbailey/dhbpapers/bcrandom.pdf}). 
 \medskip

 Compression programs like {\tt gzip} will not compact much a large file describing\footnote{For $\pi$ digits, see ftp://uiarchive.cso.uiuc.edu/pub/etext/gutenberg/etext93/pimil10.txt.} $\pi$ because they are not flexible enough pattern detectors. Their approach to simplicity (compression) is not sophisticated enough. Pi is characterized by the non-repetition of its {\em digits}, which appear randomly (as mentioned, $\pi$ is believed to be a normal number, totally democratic in its distribution of digits)! This is illustrative of the difficulty in the general compression problem. \medskip

The generation of such complex sequences out of simple rules is the subject of many studies. Chaitin would call them computable numbers (and he would remark that most numbers are in fact not computable \cite{chaitin1995}). Wolfram treats this subject in depth in his  book, ``A New Kind of Science'' \cite{wolfram2002} from the point of view of Celullar Autamata. Although Shannon Entropy was the first attempt to quantify the complexity in data,  it represents only a timid first step in the direction of algorithmic complexity. For example, the Shannon Entropy for the digits of $\pi$ is maximal and cannot capture its inherent simplicity. The order in the digits of $\pi$ is crucial to its simplicity. Shannot Entropy does not consider that and provides only a statistical analysis of complexity. \medskip

Although repetition is not obvious in the above sequence, repetition is at the heart of any compression algorithm. This is because such an algorithm will have a ``FOR'' loop: the algorithm will state that a certain number of operations needs to be repeated over and over. Thus, we can state that  {\em repetition is at the essence of compression}. An ``unpacker'', a system capable of running a program, is of course needed to realize or implement the repetition procedure.
 
\medskip
 
The brain aims to simplify (compress) and manipulate incoming information.  ``Understanding'' gives us power, speed of response, awareness of change. Our algorithms usually never reach the Kolmogorov compressive limit. This is due to noise, the fact that we only have access to partial information about our environment and also to our own modeling limitations. In what we could call ``lossy compression'', information is modeled by some program which achieves moderate compression and predictability. This is of course sufficient for most purposes. In addition, there are many aspects of the data streams that a brain may not care about, and this is another important aspect: how to distinguish between what we want to model and what we don't care about.
\medskip 

A useful image here  is that of a monkey in the jungle, enjoying a tasty banana by her lone self. Imagine an environment full of complex sounds, the daily soundtrack in tarzan-land. All these sensory inputs reach the ears of our monkey, and are acquired and processed by her brain. The monkey is relaxed. Suddenly, the jungle grows quiet. The monkey detects that a rule has been broken (the rule of ``pseudo-random sound'', perhaps) and looks up, alerted. This monkey's  brain may very well have saved her, alerting her on time about the possible presence of predators. 
 Similarly, in the same way that we try to model physical phenomena, from the micro to the macro, we must learn to model and control our environment. This is also carried out by discovering rules and simplifiers. Riding a bicycle entails the same discovery processes. We must begin by attaching a set of rules to the bicycle concept (invariants, such as shape, 2 wheels), and then learn to control the dynamics of riding a bicycle: we must encode Newton's laws in our reflexes. \medskip

Human beings, for good or ill, seem to have mastered well beyond other species the art of modeling. It is precisely this ability that puts us on top of the food chain, and in control of our trembling planet. In some ways, it is perplexing how beyond we are from our fellow travelers on Earth and how fast we have distanced ourselves from our evolutionary companions. From the point of view of the planet (as in, e.g., the Gaia Hypothesis) and our current lifestyle, this is definetely not a good thing.

\subsection{An example: compressing GPS signals}
Here is a somewhat technical example that illustrates the relationship between compression and modelling, one that has to do with with Global Navigation Satellite System (GNSS) communications. GNSS emit rather weak signals in the form of pseudorandom codes. These signals can be detected on Earth through cross-correlation with locally generated copies with the same codes, although by the time they reach the ground the emitted signals are well below the thermal noise level.
When such signals are captured, e.g.,  through 1-bit  sampling of the down-coverted data and storage into a hard drive, the resulting files appear to be basically noise. This is only partly due to the pseudo-random nature of the GPS signal. By the time it reaches the ground, the resulting signal is the sum of the original signal and a much stronger noise component. After 1 bit sampling, the voltage time series appears to be a random sequence of ones and zeros.  Yet, correlation of this information with a GPS code replica (a model), produces a non-zero result---a correlation spike appears when data and replica are aligned in time (a phenomenon which provides the basis for GPS signal detection, of course). This fact can provide the means to compress a GPS bit capture file. To see this, write
$$
S(t)=R(t)+E(t),
$$
where $S(t)$ is the bit stream, $R(t)$ is the model bit stream, and $E(t)$ is the error stream. If $R(t)$ is indeed a good model, $E(t)$ will have more zeros than ones, and can therefore be compressed using simple techniques. So even an approximate model can be very helpful for compression. \medskip

Finally, the goal may actually be useful lossy compression. That is, the user or cognitive system (in this case, a GPS receiver) may not care to keep track of the noise part. Then, essentially, S(t) will be compressed, in a lossy fashion, to a concept such as ``PRN NN with a delay of X ms and a Doppler offset of Y Hz''. 
\medskip

In general, it will be pretty hard to define what noise is, but a first attempt may be to say that noise is whatever information does not improve the chances of survival.

\subsection{On G\"odel's theorem and Algorithmic Complexity}
Algorithmic complexity provides also the backbone to study mathematical knowledge. The concept of compression applies as well to mathematical theorems and anti-theorems (statements whose negation are theorems). Once we can prove something, thus transforming a statement into a theorem, we assert we can derive it from first principles by  applying simple rules. Then, we have ``compressed'' that statement, together with all its brothers and sisters. G\"odel showed that some statements cannot be thus compressed: they cannot be reached from axioms and rules. This concept, and the complexity-compression twist, are very nicely described in Chaitin's paper in the book {\em Nature's Imagination}, ``Randomness in arithmetic and the decline and fall of reductionism in pure mathematics'' \cite{chaitin1995}, a truly marvelous piece.
\medskip

At a simpler level, something like that happens with prime and composite numbers. As we discussed above, a composite number can be compressed, in some sense, into the product of other primes. A prime number cannot: it cannot be reached through the use of smaller primes and the multiplication rule. A prime number is an island, itself a generator of new truths (new composites).
\medskip

As explained by Chaitin and discovered also by Turing, G\"odel's theorem can be stated in the context of the following fact with regards to a Formal Axiomatic System: Consistent $+$ Complete $\rightarrow$ There Exists a decision procedure.

\medskip

I include here a short glossary to interpret this statement (see \cite{chaitin1995} for a clear exposition):
\begin{itemize}
\item Formal axiomatic system: axioms plus rules to produce theorems.
\item Consistent: you cannot prove A and not-A.
\item Complete: given B, you can prove B or not-B.
\item Decision procedure: given B, an algorithm to test if B is, or is not a theorem. If system is Consistent and Complete, then there exists a decision procedure: just go through all the proofs and check that wheter  B or not-B pop out. Since by definition B or not-B are theorems, sooner or later the proof will be found (it exists).
\item Godel: there exist uncomputable statements in any consistent formal axiomatic system, making it incomplete.
\item Turing: There exist uncomputable real numbers. Hence, there cannot exist an algorithm to tell you if a program will stop (the halting problem). Hence, the halting problem is unsolvable....a Godel statement: the unsolvability of the halting problem. Also, there is no decision procedure.
\end{itemize}

Chaitin argues that mathematics should perhaps become more like physics: laws (axioms)  are added as needed, because they may be Godel statements: ``...physicists are used to the idea that when we start experimenting at a smaller scale, or with new phenomena, we may need new principles to understand and explain what's going on. Now, in spite of incompleteness mathematicians don't behave at all like physicists'' (page 42). This may be true, but  many physicists believe in the existence of a TOE (Theory of Everything) capable of generating, from a finite set of first principles, all phenomena. \medskip

I think Godel's proof of incompleteness may be a severe blow to the concept of a TOE. As such, the theory will be a mathematical theory (based on a formal axiomatic infrastructure), and it will probably be incomplete: there will be statements which cannot be handled by the theory, and which will need to be added as new axioms---just as it happens in the number theory of Godel. These may be, for instance, statements about the the masses of an infinite set of elementary particles, or the dimensions of space-time. 

\subsection{Turing machines in dynamical systems: towards absolute complexity}

\begin{table}[t!]
\label{table:computers0}
{\centering
\begin{tabular}{|c|c|}  \hline
{ \bf Turing }	 & {\bf Self-Entity} \\  \hline \hline
Box 		&  Brain  \\  \hline 
State$_o$  	&Initial Brain State \\ \hline
State$_n$    &Neuroactivations\\  \hline
Tape 	    & Universe\\ \hline
Transition Rules & Neurodynamics\\ \hline 
Tape head read & Sensors \\ \hline
Tape head write & Actuators \\ \hline 
\end{tabular}
\caption{The Turing and Brain-Universe or Self-Entity  paradigms compared. One of the key aspects is the tape-head control, which manages the information interfaces.}
}
\end{table}

The Turing machine computational paradigm is a very powerful and successful one. How does it relate to the concept of Entity, to the paradigm of a brain with sensors interfacing to the universe? And what about the dichotomy of computer, program and data? \medskip

If we think of a real physical system as a computer, and dynamics (classical or quantum) as the mechanism for computation, where is the program, where is the data,  where is the computer? In the terms of the Turing description for computation, can we identify  the tape,  the state machine and the transition rules? One cannot help but thing that these dichotomies are somewhat artificial. Physically, there are  no obvious boundaries: there is only a dynamical system.   \medskip

It is  interesting to note here that we can describe our  brain, sensor-actuator, universe, Self-Entity paradigm purely in Turing  terms. In a Turing machine we have a tape and a state machine which reads and writes on the tape. The machine has a tape head for this purpose, and it is controlled by the Turing state machine to move across,  read and write on the tape. In terms of the concepts described in these notes, the tape head acts both as a sensor and as an actuator. The tape head, instructed  by the state machine driving it, chooses which data to read (it moves across the tape) and where and how to write on it. In our language, the tape is the universe, of course. And the State Machine, with its transition rules, is the Brain or Self-Entity.   The parallel is pretty clear (see \ref{table:computers0}), yet there is something to clarify further. In our description, the brain is a modelling tool, exploring the universe and seeking to find simple rules, invariants, to compress the tape. Can we program these operations? The answer I believe is yes, we can  use a minimization paradigm. That is, we would define a cost function which evaluates the compression power of the models the program comes up with. The program is to create and test models, and shoose the shortest ones. Of course, a GA type of algorithm can be used for this, it seems pretty natural. Of course, the field of machine learning is devoted to this problem, and progress slow. The problem is very difficult. \medskip

Why did Turing machines arise in the universe dynamical system? Because they are the means to compression and therefore ``survival''. 
 The universe is a physical, complex dynamical system. Evolution and natural selection apply to the evolution of physical variables. As dynamics unfold, as time passes,  invariant structures in the data are naturally selected: that which perdures, remains. If this picture is useful, it follows that Turing machines, or self-entities, are emergent features of the dynamics, attractors in the chaotic sense of the word.   I always thought that evolution and natural selection are really mathematical laws, with a reach well beyond the biological. That which perdures, is. Here this statement refers only to information.

\medskip

Recently, authors have discussed such things as the computational power of the universe \cite{lloyd2001} or of simple oscillating systems \cite{rietman2003} and diverse physical media \cite{harding2006}. I believe the tools developed in those papers can be used to do away with the dichotomy of computer-algorithm length. That is, perhaps there is an absolute definition of KC: the shortest description of a computer plus algorithm length. Fundamental physics may provide us with the ultimate, absolute programming language, and an absolute definition of KC. \medskip

At any rate, it is important to keep in mind that we cannot focus only on the length of programs: the hardware to run programs is also important, and has to be taken into account in the evolution/natural selection analysis of Kolmogorov Complexity. \medskip

From the physical point of view, there is only one computer, the universe as a whole. As a computer, it uses physical law/dynamics as its transition rules, and its state as  the Turing tape. It is also interesting to note that the ``tape'' is the computer itself. Although this sounds a bit strange, in reality it is no different than a usual computer. A real computer is a dynamical, physical system. The programmer uploads a program, thereby setting up the ``Initial Conditions'' and ``Boundary Conditions''  of the system. When the return key is hit, the computer, as a dynamical system, unfolds. If the program is a good one, the system will reach a ``stationary  point'' (it will stop) with information of relevance to the programmer. \medskip

The universe, as a computer, is capable of simulating other computers, like the one I am using to type these notes. As a computer, the universe is definitely a universal one! \medskip

\begin{table}[t!]
\label{table:computers}
{\centering \tiny
\begin{tabular}{|c||c|c|c|c|c|c|c|}  \hline
{ \bf Turing }	& {\bf CA}  & {\bf  PC }&  {\bf Brain} & {\bf  Ocean} &  {\bf  Universe} &{\bf PDE} & {\bf Evolution} \\  \hline \hline
Box 		& Board  	&  Hardware &  Wetware & atoms &   Q-fields & paper & medium \\  \hline 
State$_o$  	& line$_o$   	& In. State &In. Brain State  &In. State   & IC& IC & In Pop\\ \hline
State$_n$ 	& line$_n$    	&Soft. State   &Neuroactivations &Ocean State  &Q-State   &  Var. Time slice & Pop$_n$\\  \hline
Tape 		& line out 	&Data Memory   &Universe &Ocean State  &QF history  & Var. history& All pop\\ \hline
Trans. Rules &CA rules &Program   &Neurodyn. &Navier-Stokes  &Q dynamics  & Equation$+$BCs& GA rules\\ \hline \hline
Tape in & line$_o$ & input data   &Sensor in & ICs  &ICs   & ICs & pop$_o$\\ \hline 
Tape out &line out &output data     & Body out  & state  & Final State  & Final State &pop$_n$\\ \hline 
\end{tabular}
\caption{Different computational paradigms and their relation to the Turing Machine. The Turing machine, a Celullar Automaton (CA), a personal computer (PC), a brain, the ocean, the Universe, a partial differential equation (PDE), a Genetic Algorithm or Evolution. }
}
\end{table}

It is also noteworthy that the ``computer embedding'' structure is actually implemented technically. A PC is a Turing Machine in the physical sense, using the laws of physics. It is used to emulate or ``embed'' another Turing machine, at the machine language level. This Turing machine is used to embed yet another one, the operating system...and son on. We thus have layers and layers of computers...the universe, your PC, the OS, your program, finally.\medskip

Perhaps a better way to look at things is to say that the universe, or any other dynamical system, provides the substrate or infrastructure  for the emergence of Turing machines. The universe itself is not computing anything at all. The universe is not a computer. Computers are emerging properties of the universe. But can computers not simulate the universe itself?

\subsection{Using Rule 110 for GA based compression}
\label{CA}
Here we discuss an approach to the generic compression problem---finding a program to effectively compress a dataset. In general, one could say that a computer is needed and some sort of programming language space to search the solution in. I consider here a Cellular Automaton (CA). One can think of a CA as a dynamic system described by an infinite set of stacked arrays in which we start from a given array with discrete values (e.g., 1 and 0) (the initial condition) and an iterative rule for dynamics. The CA is defined then by the rule that describes the generation of the next array. The rules can be many. If we focus on only nearest neighbor rules, and a binary valued system, there are only 256 rules $2^{2\cdot 2\cdot 2}$. The reader is directed to \cite{wolfram2002} for more on CAs. What I would like to disccuss here is the idea that based on the fact that the CA using Rule 110 (see Wolfram NKS \cite{wolfram2002}) is a Universal Turing machine\footnote{See http://plato.stanford.edu/entries/turing-machine/ for a definition of Turing machine.}. That is, give the right inputs, it will emulate any Turing machine (computer and program). So we have a computer (a CA running rule 110), and a rather simple description of the programming language space (the initial condition).

\medskip

\begin{figure}[t!]
\label{boxtest}
\includegraphics[width=6cm]{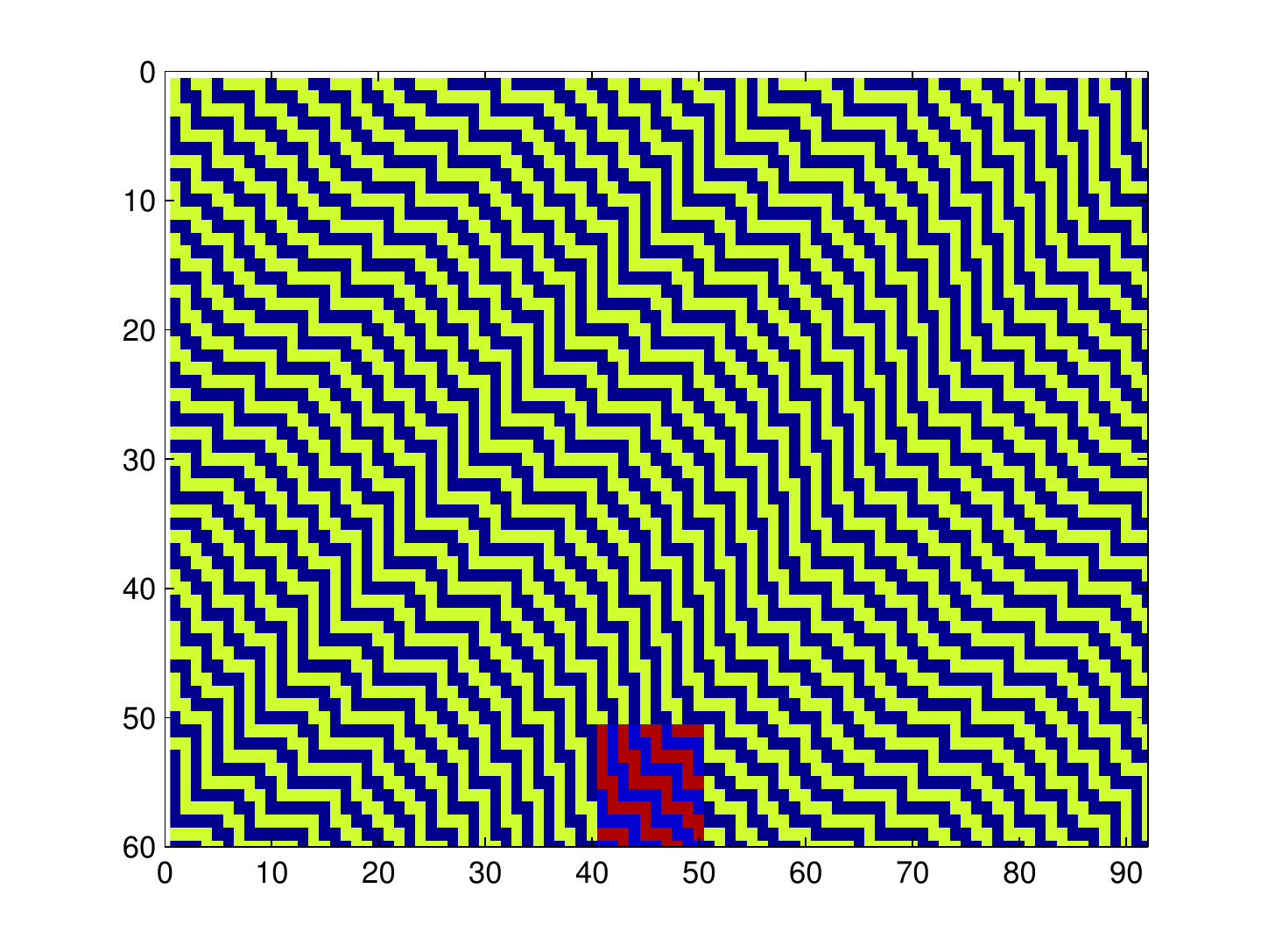}
\includegraphics[width=2cm]{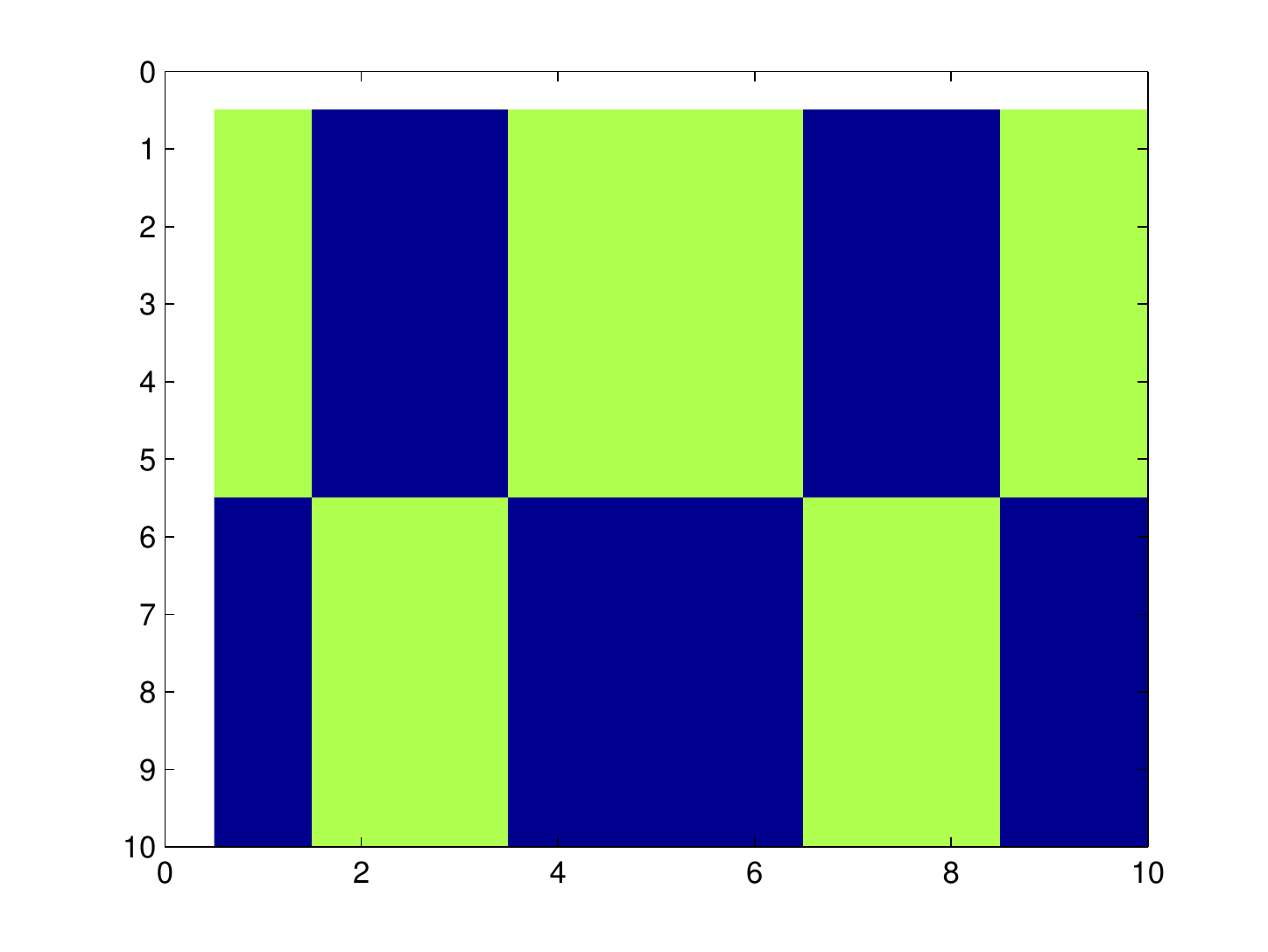}
\includegraphics[width=6cm]{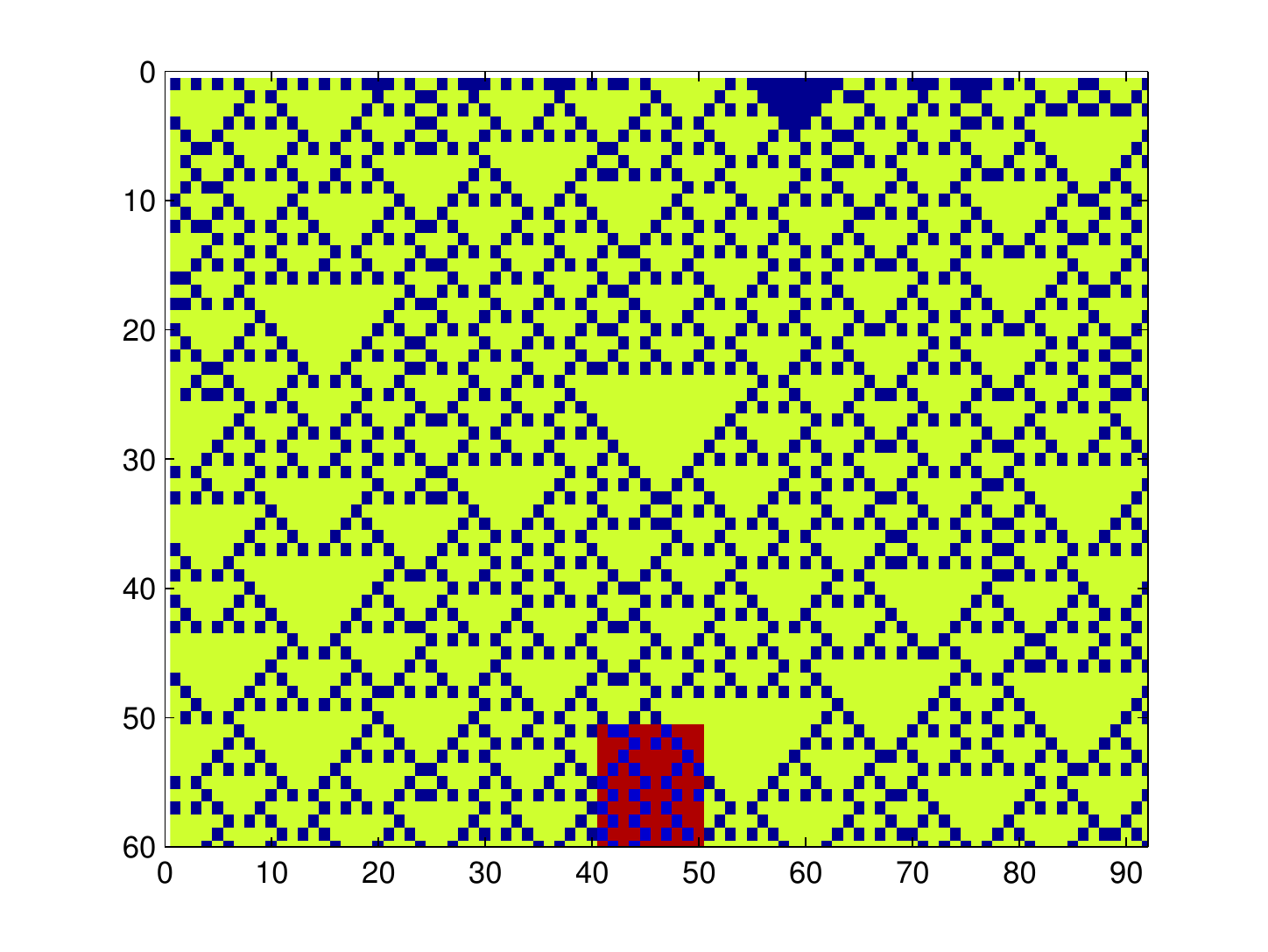}\\
\includegraphics[width=7.5cm]{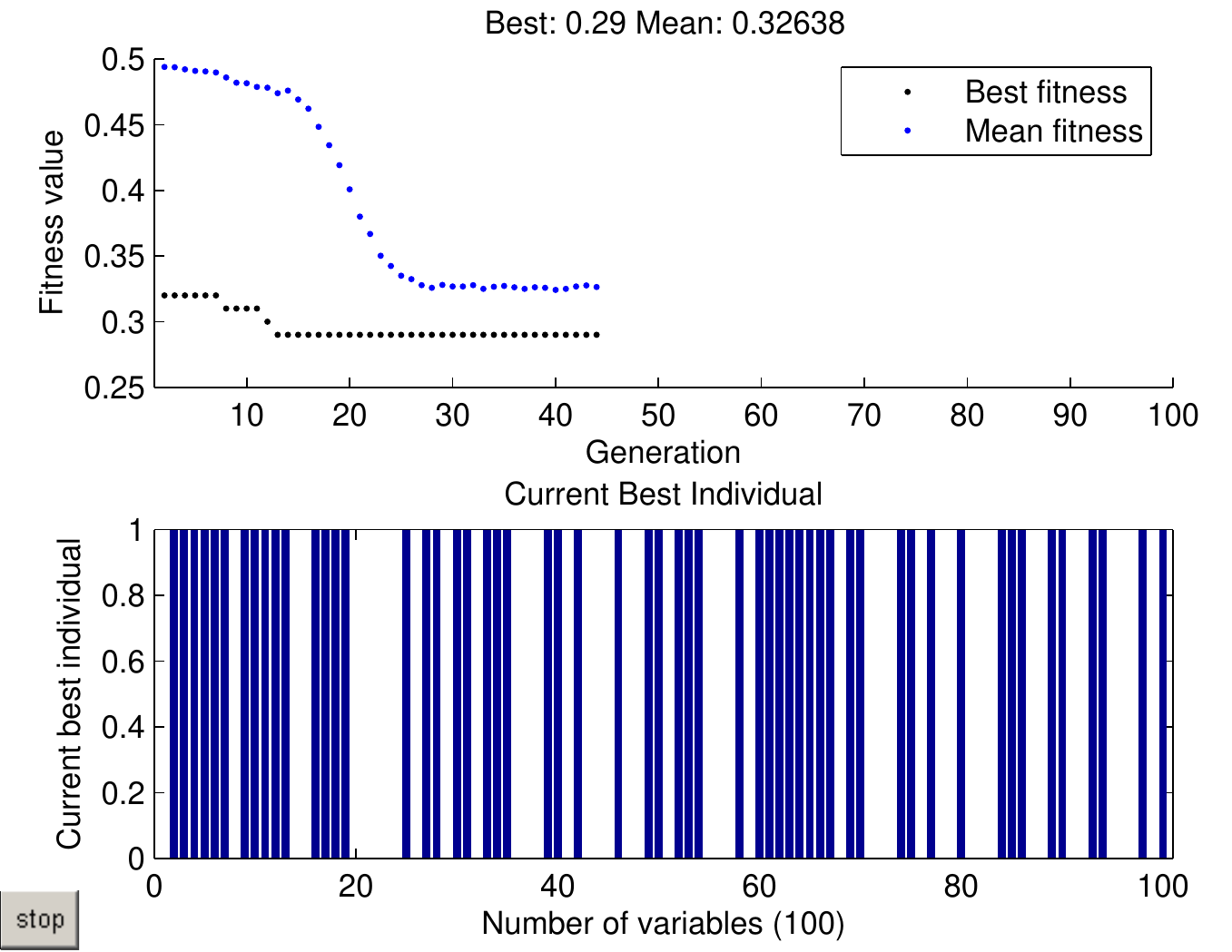}\includegraphics[width=7.5cm]{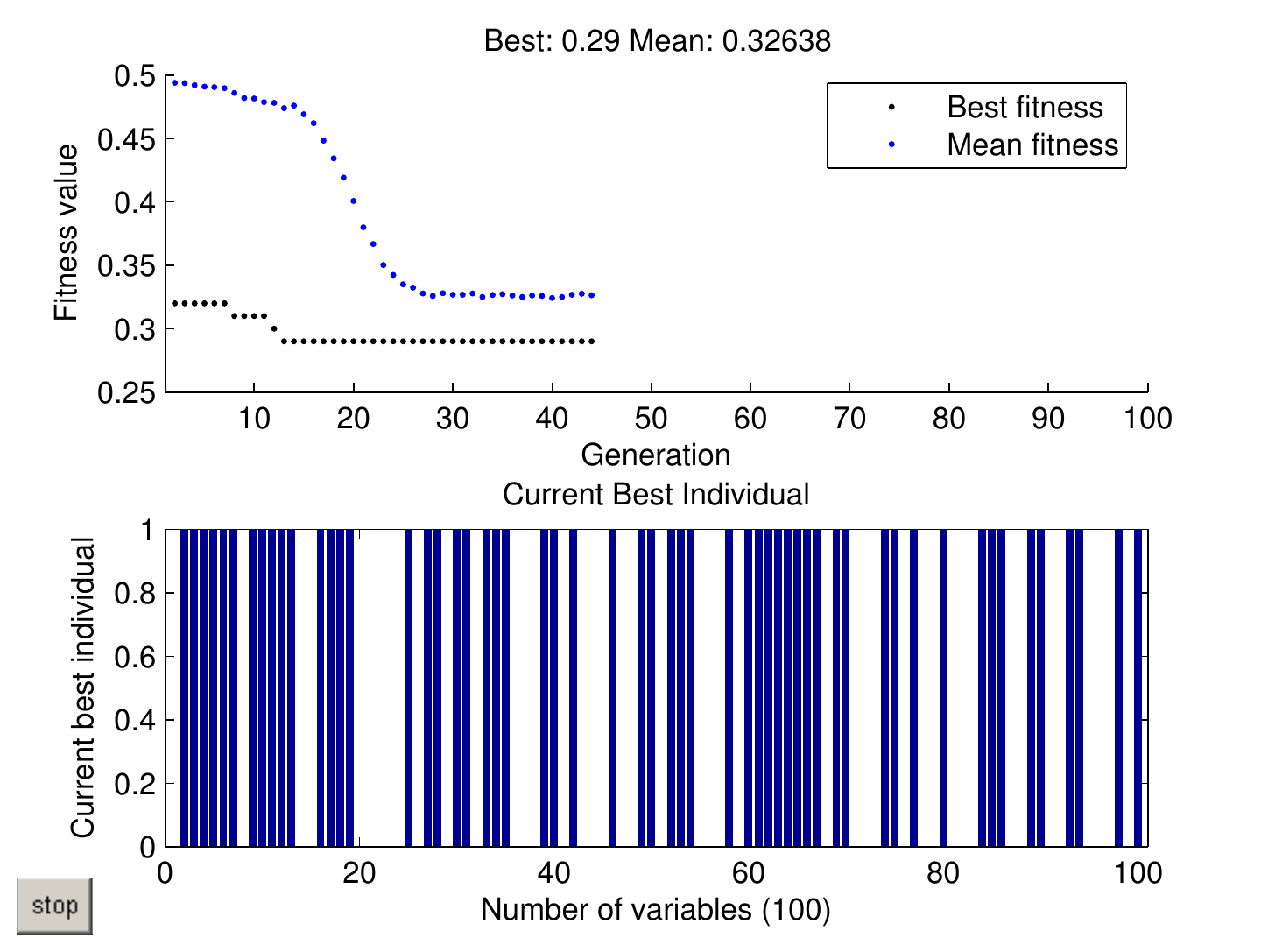}

\caption{Test solution with a box fit to target (top middle). The solution DNA is composed of 100 variables. The first 8 define the CA rule, while the rest provide the initial conditions. Top left, with any CA rule, and on the right with a fix for rule 110.}
\end{figure}

\begin{figure}[t!]
\label{test5}
\includegraphics[width=6cm]{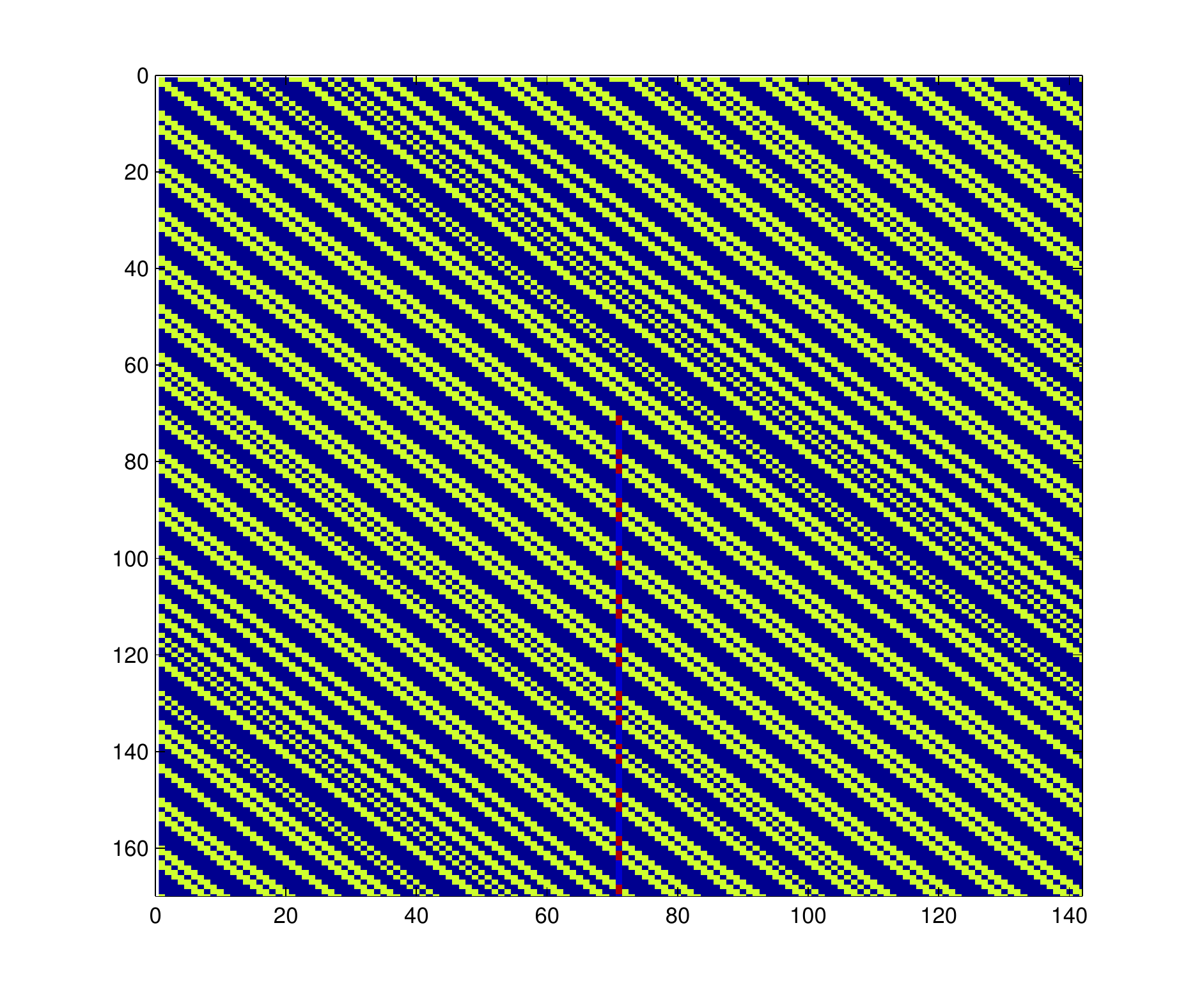}
\includegraphics[width=2cm]{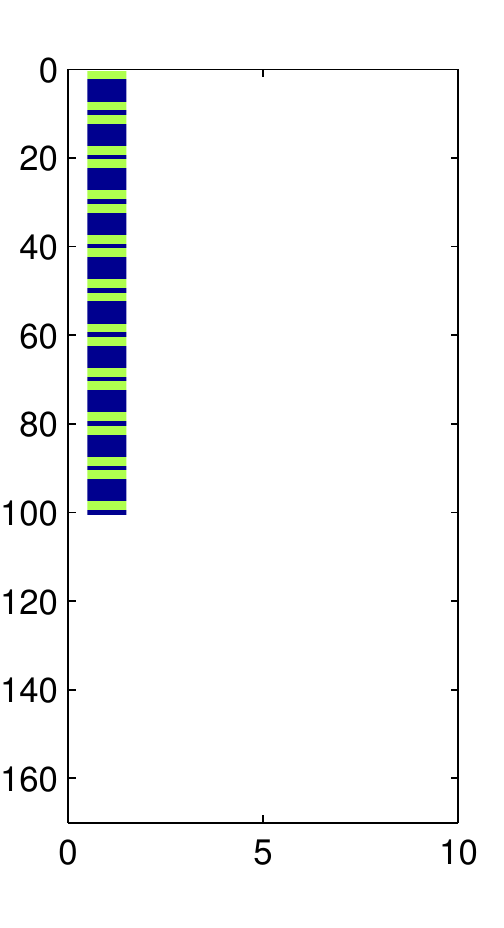}
\includegraphics[width=6cm]{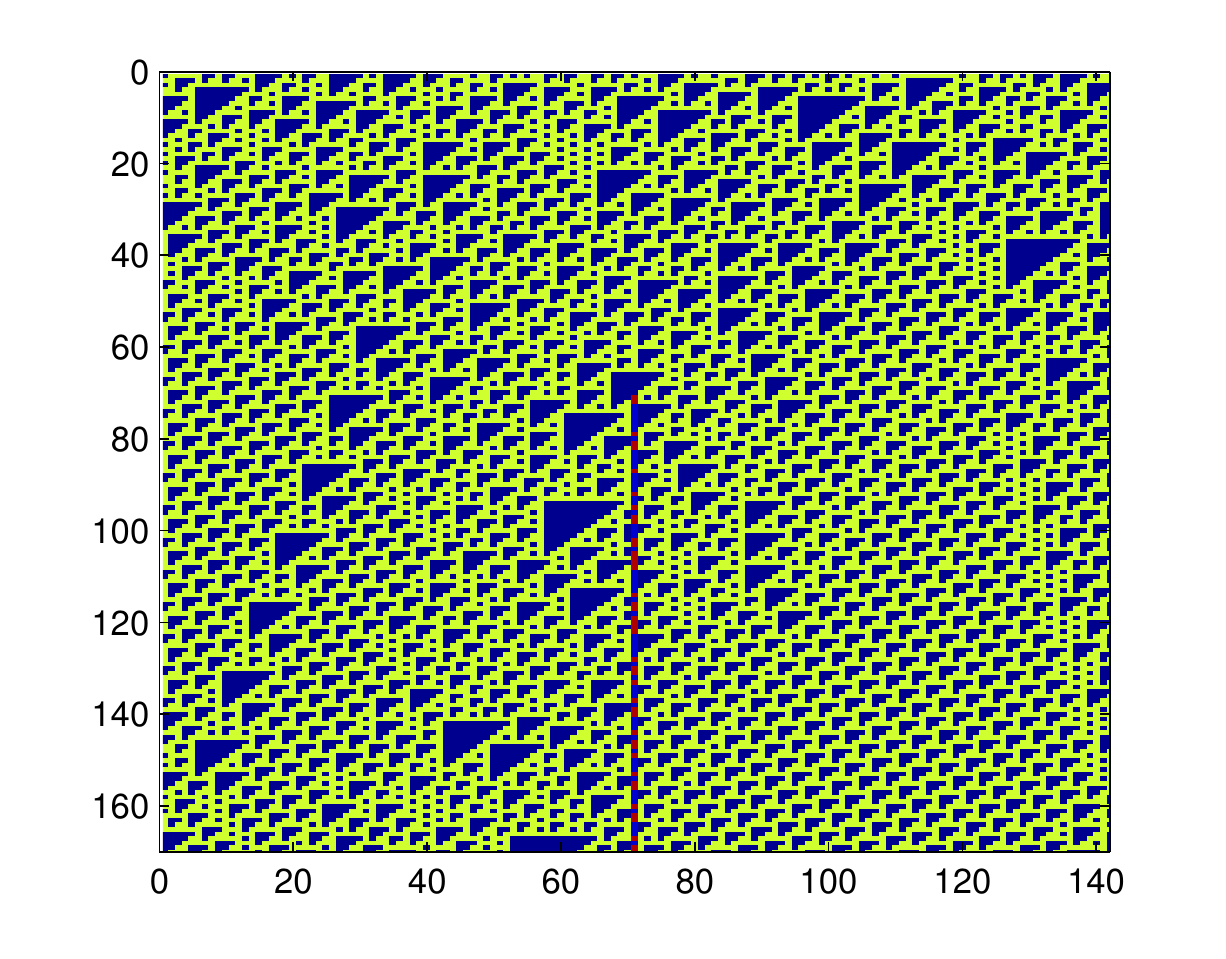}\\
\includegraphics[width=7.5cm]{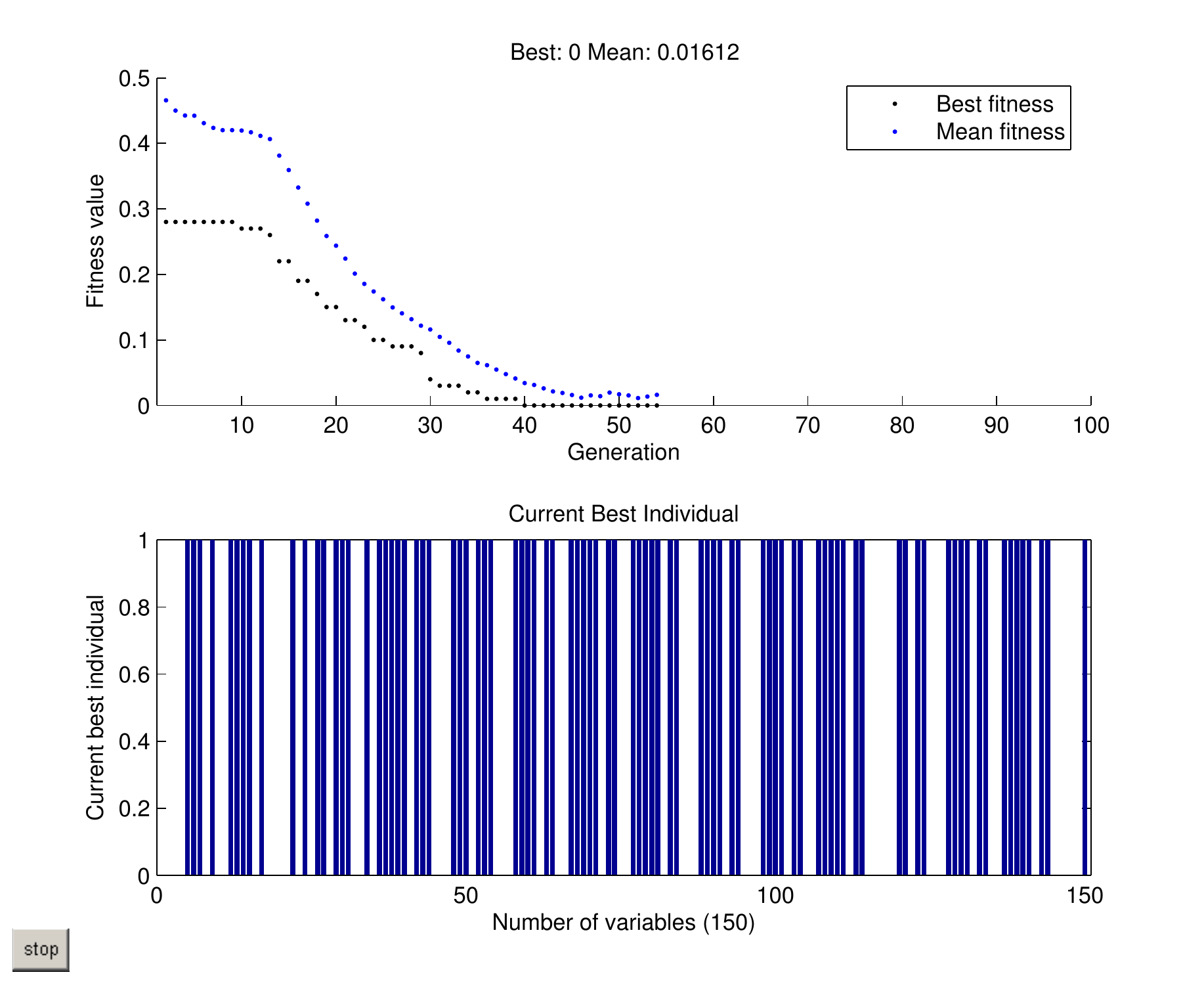}\includegraphics[width=7.5cm]{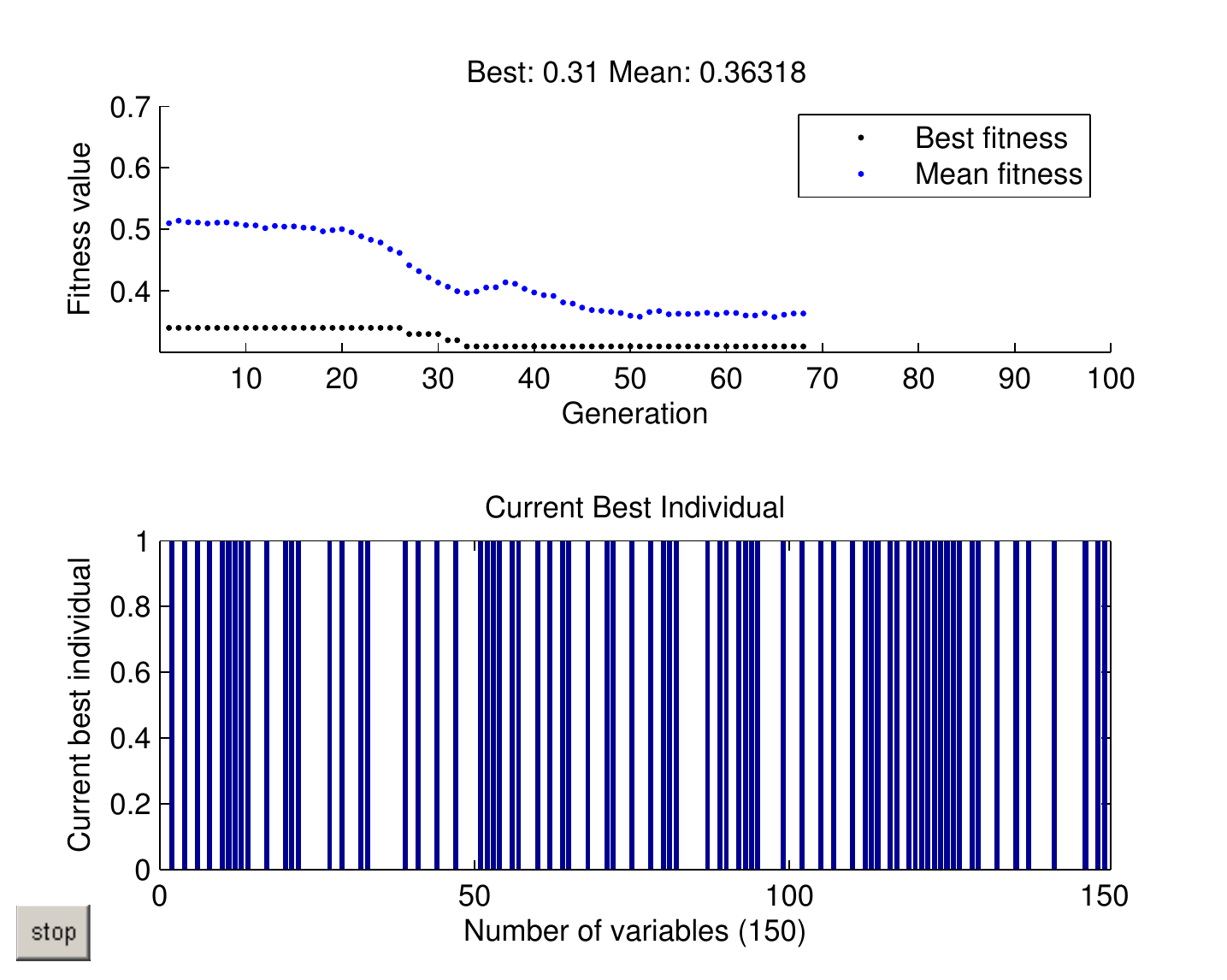}
\caption{Test solution with a line fit (time series) to target (top middle). The target is a repetition of the basic pattern [1 1 0 0 0 0 0 1 1 0 ].  The solution DNA is composed of 150 variables. The first 8 define the CA rule, while the rest provide the initial conditions. Top left, with any CA rule (112 is selected), and on the right with a fix for rule 110.}
\end{figure}

 Now, CA programming is rather amenable to a search based on Genetic Algorithm (GA) implementation. A genetic algorithm is a procedure emulating (simulating) natural selection to find solutions to a minimization problem. The solutions are born, reproduce (sexually) and die (natural selection).  If the solution space can be described by a discrete multidimensional space, then it is easy to code the algorithm. The array describing the solution candidates is used as DNA in evolution. Coding is the hardest part, because it should be done to best exploit the GA searching procedure. What I have implemented is a  GA to evolve the right initial conditions (the program) to have rule 110 emulate a set of data. The algorithm can also search in a more general rule space if desired. I would like to use Taitin's problem (discussed below) as a test case. So far, I can show basic results using rather simple patterns  (see Figures \ref{boxtest} and  \ref{test5}). The results are encouraging but also illustrative of the fact that searching in the space of programs is not trivial!  In connection with this, it would also be interesting to see if a CA approach to a Pattern Lock Loop machine would be useful.

\subsection{Kolmogorov Complexity and the Value of Scientific Theories}
There are two ways, in my opinion, of furthering scientific knowledge. One is by developing theories that can best match the observations one is trying to ``explain''. This is the standard way that has been used to assess the value of a theory. For instance, General Relativity is a better theory than Newton's because it matches in a better way the observed data and it can actually predict correctly new phenomena or otherwise poorly unexplained observations (e.g., the perihelion of Mercury, or the existence of black holes). However, I would like argue here in the context the of Kolmogorov Complexity  that a theory that leads to equal predictions but using a ``shorter program'' is also an instance of scientific progress. If we can produce the same data stream with a simpler model, we have advanced our science. At this point, the reader should not be surprised by this statement! \medskip

Thus, a theory that ``pares down'' concepts like ``time'', ``aether'', and ``scale'', and leads to the same predictions as the older, more complex theories, represents advance. Usually, and perhaps surprisingly, such simpler  theories will not only reproduce the old observed facts but also lead to new results and predictions (General Relativity is again a good example). This is just a way to rephrase Occam's Razor\footnote{Ssee http://pespmc1.vub.ac.be/OCCAMRAZ.html for more details.} 
\begin{quote} {\em 
One should not increase, beyond what is necessary, the number of entities required to explain anything.}
\end{quote}
The better perfomance, or existence,  of simpler theories could indicate that reality itself ``runs'' on simple theries. Searching for a simpler theory would then get us closer to the target, with better performance. How simple is really reality? What is the algorithmic complexity of reality?

\clearpage

\section{THE BRAIN}
Part of the motivation for this paper is the realization that we must put the brain at center stage if we are going to advance in fundamental science (physics if you ask me). We do not really know what is ``outside'' our brains. All we know is what we can interpret and model using our sensors and what they tell us about the outside world. In fact, for all we know, there may not be an outside world at all. But we suspect there must be a unifying source somewhere, however, as the inputs provided by our different sensors reveal coherence. This is not a proof of anything, of course. Given enough time we could find any finite sequence we want in a randomly generated stream (as it is conjectured for the digits of numbers such as $\pi$. Come to think of it, this thought is not so distant from present concepts such as the Cosmic Landscape (see below for a discussion of Susskind's Cosmic Landscape \cite{susskind2006}). But I think most of us expect to find simple explanations at the end of the line. \medskip

\medskip

The brain receives input, an information multi-channel (temporal)  stream. In order to gain an edge, it must try to implement a sufficiently fast predictive model of this information. The multi-channel aspect is rather important. Scientists trying to fool our senses in virtual reality environments (Presence researchers) know that synchronization, coherence,  is a key aspect in a successful charade.
\medskip

The model in our brain is constantly evaluating its performance against reality. When there is a mismatch, the brain produces an Alert event. The model fails and this gives rise to an alert signal. The goal of learning is to produce as few alert events as possible. Perhaps this provides the ingredients for a suitable definition of  pain. Pain may be equated to poor modeling and hence, painful surprises.
\medskip

In complexity terminology, the input stream has a given complexity which can be quantified. The brain develops a model that will, in general, fail to reach the compression limit  (due to hardware limitations, for instance, or simply for algorithmic search limitations). Model minus data will not be zero. The output ``alert'' stream carries the uncaptured complexity.

\medskip
\subsection{The closed Loop: self-modeling}
Part of the model of the universe that our brain seeks must take into account the model itself: the brain is not an input-only system, it produces output, and this affects the Universe (i.e., the input).  There is something very mysterious about this self-coupling, and it may be directly related to consciousness or at least self-awareness (consciousness is probably a more primitive feature). In particular, our brains decide which input streams to select and focus on (e.g., by moving or looking in a different direction, by selecting and following a particular conversation in a a party, or by changing a tv channel\footnote{If you still watch that thing. }). Modern views of AI state that ``embodiment'' is a fundamental aspect of cognitive systems. This view is easily interpreted here by stating that the cognitive agent needs to be able to act in the information environment and regulate the input information stream. That is, the agent can control, to some extent, the information interfaces (sensors), and actuators. We will see again this concept in a re-analysis of the Turing machine concept below. 

\medskip 
 
In our concept of  the Universe we must include our immediate universe, our limbs and body.  We also have to model our inertias, our latencies, our strengths and limitations. Walking is a very complex act! The propioceptive system is a key element in this self-monitoring loop, and it is probably tightly linked to self-awareness. 
\medskip

In summary, we have to model both the Universe and our impact on the Universe (something which may be the real meaning of ``us''). What we do affects us, and Reality kicks back \cite{Deutsch}. This may be the key to the self-notion (self-awareness vs. awareness, which is a more primitive notion). \medskip

The interaction between self and universe can be conceptualized like the interaction between two programs, one for the self, the other for the universe, where at least one seeks to model the other. Of course, this is just an approximation. In principle, there is only one program running: the universe.

\subsection{Pain and  Consciousness}
Why do we feel pain and pleasure? We can easily argue that this is done to ensure our survival, of course. Pain stops us from doing things that may harm us, or makes us move into more favorable situations (hunger, thirst, sex). We are rewarded with negative pain---pleasure. But  more specifically, I would like to suggest that Pain and its negation (Pleasure) are also {\em the} fundamental phenomena that drive us to model the Universe more efficiently. Pleasure is the reward to a good, simple  model. When we have the right models we improve our chances of getting where we want. And pain is the punishment for a bad model. A scientist feels pleasure when discovering a good theory, and pain all along the way. We feel pleasure when we ``get'' a joke, when we ``get'' a complex song. In fact, pleasure and pain are the sole drivers of action, and they seem to be truly fundamental for animal behavior. 
\medskip

One can also think of pain and pleasure and emotions in general, as the ``firmware'' in the brain that defines the goal functions for cognition. As firmware, it is something we cannot easily reprogram, and it resides deep in our brain.

\subsection{Sleep and Memory}
 
Why do we need to sleep? For several reasons, most likely. Although many explanations have been put forward this is as of yet an unanswered question. Restoration, repair and learning have all been proposed \cite{maquet2000, sejnowski2000, huber2004}. In my view, the last case is the most convincing one, i.e., the simplest theory to explain the facts,  or at least a very interesting possibility, and I will analyze it with the conceptual tools developed here. The idea I would like to propose is that during the day we gather more or less raw information and put it into a temporary buffer. This buffer is adapted and optimized for the amount of information we gather during our waking hours, using the time-off at night strategy we have evolved. 
During night, sensory inputs are shut down (a hallmark of sleep) and the accumulated data is processed: the apparent complexity of the ingested information is processed and reduced as much as possible, to translate it into rules, or short algorithms to be hardwired. The buffer is then mostly emptied. The discovered  rules (they are no doubt approximate rules) are better kept. They support our long-term memories and long-term models.  Thus, during the day we ``exercise'' our models (put them to work) and gather further information. During the night, we recess into our self-universe, shutting off sensors and actuators, and we indulge into model development.\medskip
 
Of course,  finding regularities (rules) in data economizes the process of memorization. As we discussed above, instead of memorizing the sequence for $\pi$ (which is infinite), just store the algorithm, it is much more economic. Compressed memories are somehow less fragile. But compression and model building are equivalent processes, and the result of a smaller file size is not the only advantadge. A model is a much more practical implementation of knowledge than raw data. Control systems for things like, e.g., a body confronted with the task of snowboarding, require efficiency and speed. Relying on old visual and propioceptive raw data to decide the strategy for the next bump won't do it. A model, a set of rules to decide the next move is a much more efficient approach. And implicit in such a model lies the body model and a good portion of Newton's laws. I remember dreaming, as a young boy, the relived experience of going down a bumby slope after the first days of the season. Just as if my body was making an effort to get up to speed---learning in my sleep.   
 \medskip 

I just read something on model building in a news feature and related article in Science (Nov 2006). In this case, the cognitive agents were robots programmed to modify models of their own bodies (and, extrapolating, their environment) from past experience and in this way improve locomotive performance \cite{Adami2006, Bongard}. In particular, the cycle would include the design of specially directed experiments to select among competing models. This is achieved through ``actuation-sensation'' interaction with the environment, of course. The methodology is an in the scientific method, where data suggest models, which are then tested via dedicated experiments. Is this what we do when we sleep? Model building and simulated testing? \medskip 

Another interesting piece of information comes from the work reported by Massimini et al. \cite{massimini2005}, in which a link is made between the phenomenon of consciusness and brain connectivity. It is shown that in NREM sleep brain connectivity is decreased as compared to a wakeful state. This indicates that the brain ``decouples'' during NREM sleep, and that consciussnes is linked to integratory processes in the thalamocortical system. If learning occurs during NREM sleep, can we argue that such decoupling is needed? Is the brain fine tuning subsystems, and then trying them out during REM sleep? 
Although this is a question we cannot answer at the moment, let me point out that some work I have been doing with other colleagues may provide useful tools to further investigate brain connectivity during sleep \cite{ruffini2005}. \medskip

In this work we analyze the complex networks associated with brain electrical activity. Multichannel EEG measurements were first processed to obtain 3D voxel activations using the tomographic algorithm LORETA. Then, we computed the correlation of the current intensity activation between voxel pairs  to produce a voxel cross-correlation coefficient matrix. Using several correlation thresholds, the cross-correlation matrix was transformed into a network connectivity matrix and analyzed. To study a specific example, we selected data from an earlier experiment focusing on the MMN brain wave. The resulting analysis highlighted significant differences between the spatial activations associated with Standard and Deviant tones, with interesting physiological implications. When compared to random data networks, physiological networks are more connected, with longer links and shorter path lengths. Furthermore, as compared to the Deviant case, Standard data networks are more connected, with longer links and shorter path lengths--i.e., with a stronger ``small worlds'' character. The comparison between both networks shows that areas known to be activated in the MMN wave are connected. In particular, the analysis supports the idea that supra-temporal and inferior frontal data work together in the processing of the differences between sounds by highlighting an increased connectivity in the response to a novel sound. This work was a first step into an interesting direction, I think, because it is clear that the brain is a complex network and complex network theory has today much to offer \cite{barabasi}. 
\medskip

A strange consequence of the  hypothesis that sleep is used for model building based on daytime data acquisition as discussed above is that our brain buffers  would be different if we lived in a planet with a 40 hour cycle, or if we were surrounded and driven by more complex phenomena. 
However, this idea is probably  testable. The more complex a task you have to learn, the more you will have to sleep. To some extent, overdoing it, burning the candle, allows us to put more in our buffer, which is essential to discover rules about complex phenomena.
 \medskip

And following this line of thought we can also ask, Why do we dream? Dreaming is perhaps related to the firing off of our memories in a random way, in the process of looking for algorithms. Stephen LaBerge defines the difference between awake and dream-like states of consciousness based on sensory input. Dreaming is like being awake without the constraints and realignment from sensory input. During waking hours, sensory input is constantly used to realign our models. This may seem to imply that our models are too unstable, but I believe it is important that this be so: we need to learn new things, and retain flexibility in our model searching.  \medskip

Dreaming may be the means to test our models in a safe way. Using stored data from past experiences, the dreamer tests his/her new algorithm for riding a bycicle, say, or for skiing. This provides a safe testbed. From this point of view, dreaming is not the most fundamental block  necessary for learning, rather one of the last steps.  Recent research on the importance of REM and other stages to procedural learning shows the relevance of REM is unclear, while the deeper sleep stages seem to play the key role.
\medskip

In its simplest form, memory is a tool to store raw information. We would therefore expect short term memory to represent data in a less processed manner. After this data is processed and simplified, it is encoded as long term memory.  The prediction we make is therefore that short term memories have a higher apparent Kolmogorov complexity, while long term ones are more compressed.
\medskip 

As as species, or a group of brains, humans underwent a  revolution  with the invention of speech. Speech allowed us to share the burden of memory and processing. Memories could be passed from generation to generation via word-of-mouth---clearly an advance, although certainly error-prone!  The subsequent invention of writing allowed humans to store in more permanent form both data and models. The transition to modern times, with the provision of an enormous power to access universal data and models will certainly have a profound impact. Thanks to developments such as the Internet and Google, it is unlikely that the old emphasis on blind memorization  will survive for long, as we now have very fast and  accessible exterior memory buffers. Emphasis will shift to data handling and processing. The true experts of the future will be masters at data-interfacing and model making, more than at storing facts. The universe outside our brains is better becoming our Turing tape, supporting our computation and making us better Cyborgs \cite{cyborg}. An alternative is science fiction's memory chip, or a direct brain interface to an external database. These technologies, far fetched as they may now sound, are surely in the horizon. 

\subsection{Neuroscience as part of the program to understanding Science and the Universe}
 
The title of this section refers to my understanding that, although physics, traditionally the most fundamental science (i.e, seeking to understand all aspects of objective reality), and neuroscience, a scientific endeavour focusing on the workings of brains and hoping to relate them to subjective phenomena, have been separate fields, the division is artificial and a better theory will have to treat them as part of the same endeavor, in a unified way. \medskip 

One aspect that needs less in terms of an explanation is rather obvious. Neuroscience has now entered the realm of quantitative science, thanks to the availability of sensors (e.g., multi-channel EEG, fMRI, direct chip to neuron interfaces, etc.) and computational resources powerful enough to attack the problems of data processing and model making. It is a new Era for neuroscience, and I expect great advances in the following decades. Similarly, and although physicists  has been able to work quantitatively now for centuries, the advent of computers has opened new lines of work, allowing them to model phenomena using tools beyond analytically tractable differential/integral equations.\medskip

As briefly discussed before, there are mechanisms in place in the brain that are well adapted to modeling patterns and detecting change. Some of the exciting experimental work carried out today  (which is what sparked my interest in compression) involves studying the response of the brain to  sound input sequences. It is possible today, with accurate timing to send some stimulus to the brain (visual, auditory, haptic) and then measure the response as seen in EEG. Typically, it is necessary to average the response over several hundreds or thousands of such events in order to filter out the noise (the brain seems to make a lot of noise to the untrained ear) and observe what is called an Event Related Response (ERP). In Mismatch Negativity experiments (MMN), which are a further specialization of ERPs,  auditory {\em patterns} are generated and the response of the brain is measured using an EEG recording system (see \cite{naatanen1978, grau1998}). The experiments can then measure the change in response when the patterns are broken. The phenomenon of MMN basically illustrates the fact that the brain, at some very primitive level, is able to learn patterns.  We can also measure how many iterations are needed before the pattern is assimilated, and we can try to understand the type of pattern that the brain's mechanisms involved are best at ``understanding''.  In this context, the brain acts like a ``Pattern Lock Loop'' system.  \medskip

At Starlab we carry out research with ERP/MMN for different applications. An interesting (very) preliminary result of our work together with UBNL is that simpler patterns appear to be more easily learned. The way this is observed is by comparing the response of the brain to the perfect sequence. When there is a {\em deviant} in the sequence, the EEG records show a change with respect to the {\em standard} response. This illustrates the fact that the brain is acting like on-the-fly modelling tool. What makes this especially interesting is the fact that the level of response to the deviant appears to be related to the complexity of the sequence. In the first experiment three sequences were tested:
\begin{verbatim}
1: ABABAB ABABAB ABABAB .... 
2: AABABA AABABA AABABA ....
3: BAABAB BAABAB BAABAB ...
\end{verbatim}
The reader may notice that the sequences have been ordered according to complexity (which one is the easiest to memorize?). Even a  simple program for compression, like {\tt gzip},  
is more efficient at compression with the first, then the second, then the third sequences.  The first sequence is efficiently described by the AB subunit and the space (common to all). To describe the second one, we need A twice  and then BA twice plus the space. The last sequence requires also two subsequences, BA and AB, the first one repeated once, the second twice.  \medskip

Our results are rather preliminary and  we have not investigated yet is the number of iterations needed for the brain to learn and assimilate the pattern. It is worth mentioning here that present understanding seems to point out that simple brain mechanisms are responsible for learning these sound sequences. The subjects in the experiments are not paying attention to the sounds, they are told to read or undertake other high level activities. In fact, depending on the severity of their condition, patients in a coma are also capable of assimilating the patterns and detecting change. There appear to be rather low level ``pattern lock loop'' mechanisms in our brain. Are these the building blocks of reality? They certainly occur at the very first processing stages, and drastically reduce the data influx passed on to higher processing levels.
\medskip

How can we describe the algorithmic complexity of these sequences? In order to do that we will have to chose a programming language. We will use a rather simple one. It will contain the simbols A, B, S (space) and R$_n()$ ({\em repeat n-times}, with 0-times meaning ``forever''). For instance

\begin{verbatim}
1: ABABAB ABABAB ABABAB ....= R_0(R_3(AB)S)
2: AABABA AABABA AABABA ....= R_0(R_2(A)R_2(BA)S)
3: BAABAB BAABAB BAABAB ....= R_0(R_1(BA)R_2(AB)S)
\end{verbatim}
We can see that the length of the programs grows as expected. Is this how the human brain does this encoding? How can we test it?

\subsection{Music and dripping faucets}
Why do we like music? This is a very interesting question. I think (and this has been said before) that there is a profound relationship between music and mathematics. As a species we seem to be quite generally and cross-culturally attached to the generation of organized sound (a working definition for music). We can frame this question in the present context of complexity and answer that we like musing because music is pattern, and we are addicted to pattern catching.  This is the reason why difficult (good) music takes a few rounds of listening before we ``get'' it (read that as ``model it''). Could MMN and music be closely related? Perhaps the same could be said of humor. When we smile and feel pleasure on getting a joke, we may be experience a litte Eureka moment in which the joke data stream is simplified by the joke meaning. \medskip

Why do we hate the sound of a dripping faucet? Because the timing of the drops is chaotic, we can't find a rule for it, yet we are teased by a myriad of possible patterns. This process of never ending pattern search  becomes physically painful (recall that we hypothesized above that pain is associated with mismodelling). It is probably exhausting, and we could think of measuring this effort using a tool like fMRI (which is used to measure metabolic activity).  \medskip

In summary, when we listen to a piece of music, we are probably engaging a ``tracking'' program in our brain, with an associated model. Listening to a  good song is an active process. In a sense, we are unwrapping a model in real time and testing against the song. It is a process inverse to model building. Perhaps there are a few parameters (tempo, transposition, instruments) that we need to adjust to sustain the tracker, as in a Kalman filter. In doing that, we are simulating reality (the song) and benchmarking it with the outside reality (a good book on this subject is ``This is your brain on music'', by D. J. Levitin \cite{levitin2006}).


 
 

\subsection{Handling information}
Why is it  said that men and women think differently ... or physicists and
fashion designers...? Because depending on our universe of interest, we need to develop different information encoding and management systems for
simplification:  information handling processes adapted to our needs and our environment. This kind of observation is part of the philosophy of the old Starlab Media Impossible Consortium (Starlab NV, also known as DF1). The Media Impossible project targeted the development of technologies for automated meaning extraction of media. Something to allow asking questions such as ``show me the portions of the game on tv with goals'', or ``show me all the movies I have in which Scarlet Johansson is wearing a green tie''.   As human beings we are concerned with specific aspects of our environment, and we develop rules to handle the part of the information environment we care about. We have concepts like ``person'' or ``shirt'', and even ``justice'' or ``cup''. These refer again to invariants, or approximate invariants in our universe. Depending on who or where we are, we have to refer to and work with  different invariants.  \medskip

A recent article in Nature \cite{wolpert2003} discusses the relationship of the propioceptory system with our physical modelling capabilities. Intuitively, it makes sense that we have internalized physical models as a result of our interaction with the universe. I am certain I knew quite a bit of physics before my first course in 6th grade, just from my own body experimentation. I would like to add  here the conjecture that women and men think differently, and in particular, have different affinity with physics and engineering,  because they have different physical and social specializations which make them interact with the Universe differently.  Their model of the universe is slightly different than men's.


\subsection{Oscillations and resonance}
Neuroscience  research, and especialy indicates that the brain can be thought of as a  ringing bell, and that oscillations and their phase relationships are tied information processing and influenced by  stimulus presentation \cite{makeig2002}. A bell will resonate, when the correct air pressure time-series pattern is coupled to it.  A sine wave is a simple pattern. The pattern is actually the periodic pressure time series, that is, the coherence and regularity of the pressure time series with respect to the Master Clock. \medskip

Can we conceive of a different computation/compression paradigm based on coupled oscillators (oscillating rings, such as those described by Rietman and Tilden \cite{rietman2003, rietman2006}, or even mechanical oscillators and beyond \cite{harding2006})? Can we make a chaotic system capable of ``tuning'' to patterns? This would seem to be closer in spirit to the concept of computation using dynamical systems. \medskip

Could we design  such a model and simulate it? This is the goal of creating a ``Pattern Lock Loop'' machine. 

\medskip

The oscillatory paradigm seems to be taking a more and more prominent role both in neuroscience and in robotics. The common link is non-linear dynamics. Attractors can provide a mechanism for dynamical based  memory storage and processing. In effect, memories can be represented or encoded by particular attractors. Biologically, the brain complex can indeed be seen as a very large set of coupled dynamic components. Bio-inspired approaches to robotics seem to be yielding interesting results.

\subsection{Presence}
Reality, according to this paper, is a model or construct in the brain of the beholder. As has been argued, among all the possible models that can account for our observations, the simplest ones are ranked higher by the brain. The only limitation to our construction of reality  is the capability of the brain to come up with good, simple models. Nevertheless, it should be clear that the enhancement or manipulation of sensorial input to our brain should therefore have a very  powerful effect. \medskip

{\em Presence Science} studies how the human brain constructs the model of reality and self {\em using replacement/augmentation of sensorial data} (VR and beyond). \medskip
 
Presence belongs to a wider class studying how cognitive systems build models of their environment and interact with it. The field was originally  inspired by the subjective experience of being ``there'' and how to achieve and modify that experience in virtual or augmented environments.  
\medskip

In more general terms, then {\em Presence} is an emerging {\bf field} focusing on understanding the cognitive experience of being (somewhere, someone, sometime, etc.)  and developing technologies to generate and augment it (being someone or something, somewhere, sometime, without physically being there). \medskip

By nature a deeply interdisciplinary field, Presence spans a wide range of subjects: from neuroscience and cognition to artificial intelligence, sensors and systems. Aided by new technologies such as virtual and augmented reality, AI, wearable displays and high-end cinemas, Presence research aims at empowering us to achieve realistic feelings and experiences when immersed in a virtual environment. The feeling of being somewhere is the result of a complex interaction of many technological, nontechnological and personal factors. A fundamental understanding of these factors will allow for construction of virtual and augmented environments with greater effectiveness.
Presence research will lead to many new technologies and enable powerful applications: communications, learning, robotics, etc, that are more affordable and usable in the workplace, at home, in school and even on the move.\medskip

Although one can detect elements of Presence in simple situations (reading a book, watching a good movie, holding an engaging  telephone call), true Presence targets full inmersion, and this requires sophisticated technologies for Virtual Reality---some of which have yet not been  invented. Note here that Presence is not the same as Virtual Reality. The field of virtual reality focuses on technology development. Presence focuses on the cognitive aspects needed for ... presence, and it shall guide technology development.\medskip

The following fundamental research pillars of Presence can be identified: Human Cognition, Machine Cognition, and brain-machine interaction. All of these are essential aspects of Presence. And they have at their center or overlap the phenomenon, the subjective feeling of some reality  with other mediated agents, human or machine.\medskip

As one important working methodology, Presence uses the benchmark\footnote{Definition from EU FP6  Presenccia Integrated Project, coordinated by Mel Slater, UPC (2005).} for  {\bf the successful replacement/augmentation of sensory data with virtual generated data}  with success defined by analyzing the response of the subject in physiological, behavioral and subjective terms in relation to a real situation (this is the concept of ``ground truth'').
But Presence can deliver much more than normal reality.
\medskip 

\medskip

As a field, Presence,  in more than a way, is the science of existence. It can also be called the science of illusion. Research in this field  aims to understand and measure our experience of being (somewhere, somehow, ...). In many ways, we have all encountered situations in which some means are used to alter our ``presence''. Two examples will suffice. The first is what I would call the experience of  ``nightlife''. Why do people party more at night (as opposed to during the day)? Because at night the photon flux from the sun is greatly reduced, and therefore so is the natural sensorial input into our brains. At nightime it is easier to artificially manipulate our sensorial input. Anyone who has experienced a disco bar knows what I am talking about. 
\medskip

A more primitive form of Presence  is illustrated perhaps by the 3D  vision drawings fashionable in the 90's, where the sensorial input to both eyes can be made ``coherent'' by slight divergence of the eyes' gazes, and thereby provide a simpler explanation for the brain. Coherence of sensorial input, understood as a simplifying agent, is an important elemente for Presence. \medskip

Another example is the ``presence'' one can feel when reading a good book. You can feel you are ``there''. Here we have a sensorial replacement by sensorial input generated by our own brains (at some level). Imagination is perhaps the best known tool to experience presence. And, of course, dreams can be extremely vivid. 

In some ways, humanity has been striving for  higher Presence: storytelling,  books, theater, cinema, caves and  now virtual reality... Reality is the ultimate technological challenge.

\medskip

Following the logic laid out in this essay, we can state that in order to echieve ``more'' Presence, simplicity in the inputs is a key aspect. We state this as an hypothesis:
\begin{quote} {\em 
Given alternative models for the sensorial input, a brain will select  the simplest one it  can construct. }
\end{quote}
For example, in the ``fake''  hand experiments  using a rubber arm, virtual, or mixed reality setup as described in \cite{ijsselsteijn2005}, the subject will tend to select the simplest explanation. In the experiment, the real forearm is hidden, and a fake one is displayed and stroked. While the fake arm is stroked, the real one is stimulated  but hidden from view. In this case, the models the experimental subject observing the fake hand and experiencing real hand stimulation would consider could be  
\begin{itemize}
\item That is simply my hand being stroked
\item That is my hand being stroked by a brush as seen through a TV
\item That is the image of my hand as seen through a TV, but stroked by a real brush
\item There is a complex set-up in place to fool me into believing that that is my hand
\end{itemize}
The complexity of the ``explanation'' of the experience increases from top to bottom. The rubber hand experiment can be fitted by the first model. Of course, the better rendition of a hand we can provide, the less noise the subject will need to deal with to accept the illusion. \medskip

Another example is provided by the ``Pinocchio'' illusion. In this set up, a blindfolded subject is made to stroke a third party nose while his/hers is  simultaneously stroked. The coherence of inputs (haptic inputs through nose, hand and propioception) supports the ``I have a long nose'' theory. A higher level explanation involving the cortex (``there is a set-up to fool me'') is more complex and disfavored: keep explanations simple.
\medskip

To summarize, we conjecture here that  the feeling of Presence, as measured by subjective or objective ways, is increased if the induced sensorial input (the input data stream), has a low complexity, i.e., it can be modelled in a simple manner by the subject's brain. Coherence in the inputs, in this sense of there being a simplifying model, is an important element to enhance Presence. Bayesian expectations are also an important aspect: explanations with a better match with past experiences are inherently simpler. 
\medskip

\clearpage

\section{LIFE}

What is Life? Who invented that? I would like to argue that Life can be analyzed in information theoretic terms as well:
\medskip 

 {\em A living being, or Entity or agent,  can be defined to be a replicating program that successfully encodes and runs a (partial) model of reality, thus increasing its chances of survival as a replicating program.}
\medskip

Thus, an {\bf Entity} in this discussion is a perduring information conformation.  For example, a prion, in an information or algorithmic  sense, is a stable information conformation (representing all prion  instances). There is only one ``prion algorithm'', and this is the perduring Entity. \medskip

The terms ``agent'' or ``entity'' are interchangeable here. The first is a more common  term in the AI community. The second highlights more the implicit subjectivity of being. \medskip

If we imagine a Celullar Automaton model of the universe, a perduring information entity is a subset of cells that more or less perdure under time evolution---e.g., we can imagine a set of cells which in time moves around the board (as in the Game of Life). \medskip

Here we face again the boundary problem: there is no natural {\bf bondary} line between a program, data and computer in the physical universe. In the same way, there is no obvious way to draw the line between ``me'' and the Universe, and the concept of Entity is already fraught with ambiguity. In the same way, the Turing machine paradigm is lacking. It appears to require an artificial boundary, not unlike the observer-universe boundary in Quantum Mechanics, which has created so many unresolved conceptual problems. \medskip

Here we encounter a recurring problem. The concept of compression relies on the concept of ``data stream''. Like water in a river, information flows from the universe to the entity. {\bf Time} is needed in this discourse---apparently.  \medskip

Time is needed for to talk about Computation. Without the concept of Time, all we are left with is the Program. But note that the Program contains already everything, the Computation is intrinsic to it. In this sense, ``Time'' is in the sequence of digits of $\pi$, each digit a tick of the clock. The algorithm freezes Time.
 \medskip

The dichotomy of Program and Computer is also associated to Time---the computer provides for time evolution, it is what in QM is called a Propagator---so this reflection may provide the way to do away with it.

\medskip

Part of the code that makes up a spider includes the rules to build a web. A spider web, in this sense, is the output of a smart program that encodes part of reality (the reality of flies, wind, rain, hunger, etc.).  A successful DNA string encodes part of reality.  The input of this DNA code is its immediate chemical environment. Its output is the proteins and other RNA strings it will make, given the input. A successful DNA molecule can also control its environment to a healthy extent. In order to do that it must encode a good model of how this environment is expected to behave. In fact, we should really think of DNA-based life in the context of a theory of interacting Turing machines---at least as a first step. \medskip

Another way of seeing this is to realize that by looking into the DNA of a living organism we can gain knowledge about the environment in which it is meant to populate. \medskip

It follows that the mutual information of the code and its environment must then be non-zero.  We can find similar ideas in \cite{Deutsch}:
\begin{quote}
{\em So living processes and virtual-reality renderings are, superficial differences aside, the same sort of process. Both involve the physical embodying of general theories about and environment. In both cases theses theories are used to realize that environment to control, interactively, not just its instantaneous appearance but also its detailed response to general stimuli. Genes embody knowledge about their niches. Everything of fundamental significance about the phenomenon of Life depends on this property, and not on replication per se.....Life is the physical embodiment of knowledge. }
\end{quote}

The Brain is therefore a natural evolutionary step in Life: to build a large organism and complex organism it is more convenient to pass from few to many sensor, to getter a better ``view'' of reality. The brain has perhaps evolved to process massive input from  sensors. \medskip

Natural selection drives the universe  towards the establishement of perduring complex entities. Why? Because to perdure,  information about the environment must be encoded in an efficient,  executable form in the perduring entity---and the environment is complex. Anything that lasts must ``know'' about its environment and be fast in its knowledge\footnote{Here is an interesting idea for a virtual experiment:  run Tierra with  a complex environment background, to demonstrate the evolution of complex programs.}.  \medskip 

This is a generalization of the usual genetic/reproductive  description of natural selection.  Thus, Natural selection is the reason behind our Kolmogorov minimization capacities. Only those entities that can model and thus have the opportunity to control their environment the best will survive. This is entirely in line with the definition of Life given above. However, Natural Selection and Evolution can be based on a biological information handling (such as DNA and genes), or on other concepts (e.g., Dawkins' memes). The key concept is that of a program---regardless of its physical instantiation of the memory and computation system.
\medskip

The high level activity we call ``Physics'' is a clear example of our brain's never ending goal of simplifying the universe (of sensory inputs). A simple example are Maxwell's equations, which provide a good model for electromagnetic phenomena (and associated data streams). In four lines (and the machinery behind) we can compress radically the information associated to electromagnetic pheonomena. Physical theries achieve their maturity when the can account for a wide range of phenomena with a few equations and axioms. \medskip

A more fundamental illustration of compression comes from the concept of invariant.  Take the concept of ``mass'' or ``charge''. What is mass? Mass, a physicist will tell you,  is simply an invariant. Mass is a rule that says
 \begin{quote}
       ``In your data, if you combine the information from this and that  in such a way, the result will not change over time or space. This quantity can be used to predict future dynamics.''
\end{quote}
In special relativity,  (invariant or rest)  mass is the invariant associated to elemetary particles by $\left( m c^2\right)^2=E^2-\left(c p\right)^2$ (where $c$ is the speed of light). That is, it embodies a rule that tells us how to combine energy and momentum measurements is a particular way, to get always the same number, no matter what the particle is doing or in which frame of reference we are measuring from. The concept of mass, at fundamental level, is a rule, a model for  what we observe. \medskip

The same applies to  charge, spin, energy, particle, etc., and even time. None of these really exist in the classical sense of the word. They are just elements in our models of what exists.  Julian Barbour states in his recent book, ``The End of Time'' \cite{Barbour0}, 
\begin{quote}
{\em What we call time---in classical physics at least---is simply a complex of rules that govern change.}
\end{quote}
According to this point of view, {\em Time is definitely not a fundamental concept}, just another rule, another gear in our model. So Time should not be a fundamental block in our Theory of Everything! The same can be said of any other physics rule: mass, space,  spin, charge, ..., none of them are fundamental reality blocks. The substrate of Reality is surely not ``material''. Materialism is a red herring.\medskip

Rules, invariances: there is simplicity hiding in the apparent chaos. Conservation laws are a special type of invariance laws: they refer to invariance under time translation. They are part of the Program in plain language. In classical physics, once the equations of motion are solved, the entire history can be succintly summarized by defining and giving a value to the invariants. This can be achieved, for example, by a clever change of coordinates that make the Hamiltonian zero.  \medskip
 
There is another interesting connection between the Kolmogorov limiter and time. Once you have the limit program, you have frozen Time. Time is related to the data stream, but once the Program has been found, time is just a crutch that falls off. If the program is not the limit program, then you need less data stream. In a way, you have less need for time.  \medskip

Once you have an invariant, Time disappears. This is an interesting connection. From this point of view, the Brain seeks to do away with Time as well. Physics is based on ``dynamical laws''. Time is part of the ``unfolding'', or decompression of the universe model. \medskip

If someday we do find the Theory Of Everything, it will read something like Maxwell's equations, something of the form\footnote{Feynman points out in his Lectures that one can vacuously write all the equations of physics in an apparently simple form: $U=0$. This is not what we mean here! Inventing notation is not a valid way for compression.}
$$
       dF=0.
$$
The goal of finding a Theory Of Everything is to find  the ultimate compression algorithm.  We will then have the algorithm to generate the universe in a faithful way.  
If we do find a final model, we will still need a big computer to unfold it into predictions. In, fact, there may not a faster way than to "run a universe" to see what phenomena the model contains. But what is the Kolmogorov Complexity of the Universe? And where is the boundary across programs? Where does the program end and the computer begin? Such distinctions are clearly artificial. From a certain point of view, Mathematics can be see as providing the computer language to write the programs of physics.
 \medskip

Of course, if the algorithm can itself be simplified, we will not have finished yet. The form of the final theory, if we find it, may be different than expected. Susskind, in his recent book titled "The Cosmic Landscape" \cite{susskind2006} addresses the issue of simplicity of Physics Laws in a direct matter. Pointing out the extreme fine tuning of physical constants and particle types for the benefit of Life (notably, the cosmological constant) he asks if we can expect to find a simpler physical model to encompass all, or if there isn't a real explanation, if a fine tuned theory to explain all the miracles does not exist. The second option would be compatible with the existence of a multiverse (megaverse as Susskind calls it) that would allow for all possible combinations of physical constants (and theories) in a multidimensional cosmic landscape. In this framework, we observe the precise and miraculous combinations of constant values that we do because we, as human beings we are part of that solution, of that point in the the cosmic landscape. This is the modern form of the Anthropic principle. That is, there are many other universes in the multiverse, but in most of those there are no human beings (or organic matter, or even stars) there to observe it. It was once hoped that String Theory would impose a complete set of constraints on all the constants and particle types, but this is not the case. There is a landscape of solutions. This landscape gives scientific credibility to the anthropic principle. This principle is in fact very much in line with the ideas in this paper, where the brain is at center stage. Anyhow, I feel it is rather premature to jump to conclusions on the Landscape and the Anthropic principle. After all, string theory is not the widely accepted way forward (although this may now be changing). Nevertheless, the landscape possibility is rather interesting. The question then is, what law defines the landscape (in this case some final form of String theory). The goal of simplicity in explanation is certainly still a valid guiding theme. These new developments  are in no way a departure from traditional scientific work.  \medskip

Here I would just like to point out something I concluded during  my thesis work \cite{mythesis}. That is, that gauge theories are just a mathematical constructs to be able to work in topologically non-trivial phase spaces. This concept seems to mirror nicely the point of view of Barbour on time. In essence, time is gauge. There is something of a mistery here: to simplify things we must enlarge first the space of discourse! This sounds like a shift of complexity from the program to the programming language (which now needs to include, e.g., ghosts in the BRST formalism). 
\medskip

\begin{figure}[t!]
\includegraphics[width=14cm]{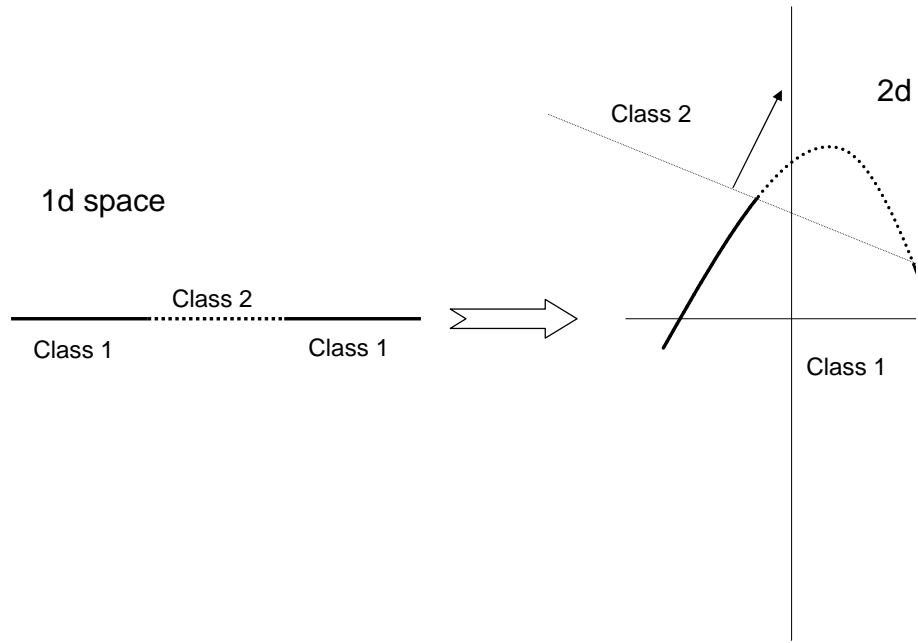}
\vspace{-5cm}

 \caption{ \label{svm} How to solve a classification problem by increasing the number of dimensions and using a non-linear map. In the 1d space case there is no ``plane'' (point) which can divide the space into two sections with the correct representation. In the 2d case, after a non-linear embedding, this is possible---a hyperplane exists that separates the two classes.}
\end{figure}

At any rate, the point is that by adding dimensions or degrees of freeedom to a problem the description turns out to be simpler (in the algorithmic sense). In reality, any calculus student knows this from the use of Lagrange multipliers to solve a constrained, minimization problem. Adding the Lagrange multiplier dimension (one per constraint) make the problem easy to solve, as opposed to first imposing the constrain and then solving the minimization problem. I think there is something deep here that appears in fields as diverse such as constrained minimization theory,  Gauge theory \cite{ruffini2002}  and the theory of Support Vector Machines \cite{cristianini2000}. In the latter case, the problem of classification using hyperplanes in some dimension is solved by a map into a higher dimensional case, where the it is possible to separate cleanly the two desired regions using a dividing (linear) plane---see Figure~\ref{svm}. Embedding and non-linearity are both needed because the topology of an easy-to-classify problem is trivial, while in the starting feature space topology of the class regions will be complex.

\subsection{The Theory of Interacting Turing Machines,  Aging and Cancer} 
Consider a fertilized egg, the fusion of sperm and ovulo.  This is a program, encoded in the DNA sequence. This program is immersed in an environment,  the chemical environment of input and outputs.   Part of the program includes a statement that says, ``when the environment is such and such, divide''.  We can then think of this as a Turing machine, interacting through the environment. An organism is a society of cells, so we can say that at the end an organism is made up of a very large number of Turing machines. 
They form and organism because the communicate efficiently and because they are somehow ``aligned'', acting coherently to  help the organism survive. As time goes by, small changes occur in each Turing machine, and these changes are not all the same. This degrades the coherence of the organism, and the organism ultimately dies. \medskip

This is perhaps why we reproduce starting all over again, from a single, coherent program, the single fertilized egg.  
In a recent paper it is argued that evolving from a single fertilised egg makes evolvability more likely \cite{wolpert2003}.  
\medskip 

The issue of aging is also related to {\bf cancer}. At some point it appears that some cells in our body rebel against the system and become autonomous, rebellius, hijacking organism resources until it ultimately fails. This is a rather mysterious event. It would seem, from a natural point of view, that the tumor and its colonies become in many ways an independent organism. That is, a tumor is a rather complex machine, capable of hijacking resources, evade safety mechanisms, etc. Such an  event would not appear to be an accident---rather the opposite, there seems to be purpose here. However, if the tumor is a new organism, a parasite of the host, how does it reproduce, how does it enter the natural selection game? It is interesting to note that recent studies link cancer and stem cells---cells not unlike the original fertilized egg (see Scientific American July 2006).


\medskip

Alternatively, cancer is a fast self-destruct mechanism to expedite demise, as we discuss next.

\subsection{Natural selection as Computation}
Why do organisms age and die? An idea I would like to consider is that this is  because this is the solution found by Nature 
(in our local universe) to make perduring entities. Here we could distinguish between entities and meta-entities.  Entities are what I called before living organisms. Meta-entities would be the pattern defining the Entity. For instance, a prion would be considered an Entity, and there can be many instances of this entity. There is only one meta-Entity, though, the pattern for a prion. \medskip

That is, having organisms die allows for more room for reproduction, mutation and genetic algorithmic improvement.
Clearly, in a finite world with finite resources, without  aging  or death, there is no space for evolution---there is no room for reproduction once the resource capacity of the habitat is exhausted. 
And if we are in competition with other evolving organisms, that is not a good strategy. So, we must die and reproduce. And the 
perduring entity is not the individual, but the individual ``stream''. Here I note that environmental stress is a cause for cancer---which is a fast route to demise. Perhaps we accelerate our ``turnover'' when the environment sends us agressive signals.

\medskip

Evolution here is to be thought of as a form of computation, the means to obtain perduring entities. \medskip 

Of course, evolution is only needed if ``perdurance'' is under stress. If we have found the solution to
perdure without biological evolution (or, at least, natural bio-evolution), aging and mortality are no longer a must.  
I think we will some day stop and eliminate them in our current ``individuality'' culture. The stream will stop, the pattern will remain.

\medskip

So, this reasoning fits well with a picture in which:
\begin{itemize}
\item The Universe is computation.
\item Entities are programs (in an abstract sense, that is, two copies of a running program are effectively the same program).
\item Programs evolve. This is also a form of computation (a Genetic Algorithm).
\item Programs that perdure, perdure and are around. Note that ``perdurance'' can result from successful reproduction (bacteria) or high resilience (diamond).
\item A program will perdure if it has captured the essence of its environment.
\item In order to perdure, sometimes it suffices for a program to evolve slowly. Other times, when the environment demands it,  it better evolve faster---competition is tight.
\item Computation cycles and their duration, at some level, are defined  by the die-mate/mutate-reproduce cycle (GA).
\item Computation (reproduction) consumes resources, and it should be avoided if possible.
\item Thus, there is an optimal ``age'' for every program which depends on the environment: a hostile environment to the ``program'', invites a faster clock (shorter lifetime, perhaps through cancer or other self-destruct mechanisms), a benign environment the opposite. Turtles and elephants have long lifespans. Bacteria the opposite.
\end{itemize}
It is worth noting that recent research (e.g., see New Scientist, April 19th 2003, p  27) indicates that organisms have a ``lifespan knob''. That is, there are molecular knobs that can alter lifespan.
\medskip 

These ideas suggest a virtual silicon experiment: let the lifespan itself be a genetic trait that can be altered, and see how a species adapts to different environments (with limited resources penalizing fast reproduction). In terms of a GA, we are saying that the cost function is changing with time (perhaps suddenly). As a result, mutants with shorter lifespans will adapt faster and take over.
\medskip 

Finally, note that the title of this section hints at the fact that there are many other forms of computation. Even in evolutionary computation, the use of ``reproductive hardware'' is not strictly necessary. Memes are another example of programs that can evolve and perdure. The use of DNA for information technology is probably just one approach among many possibilities.

\subsection{Maximum Entropy Production}	

Several papers have appeared recently that point out to a very interesting link between complexity and non-equilibrium thermodynamics. It has been shown that non-equilibrium systems such as the Earth (driven by the Sun) settle into a state of maximal entropy production given the available limits (constraints). In statistic mechanics terms, such states have the highest probability. There could be an interesting link between MEP and KC in the context of Life. If systems such as the Earth-Sun seek MEP, then Life may search for powerful analysis tools of the hyper entropic data streams. The universe, our universe is a complex place.

\clearpage 

\section{PHYSICS}
\subsection{Time and Kolmogorov Complexity}

What is Time? Julian Barbour calls it The Great Simplifier, in reference to Ephemeris Time. I would like to argue that our brains have found Time precisely because it greatly simplifies our understanding of the universe. We find, in Barbour's book, an interesting quote of Mach that says it all:
\begin{quote}
... time is an abstraction, at which we arrive by means of the changes of things.
\end{quote}
In Barbour's book we find a very clear exposition of the so-called Tait's problem. The problem is as follows (the reader is invited to \cite{Barbour0} for details). Suppose we generate on a computer a set of numbers (triads), representing the separation between 3 particles in a free motion, non-interacting: 
\begin{verbatim}
      ......        .....       ......
      7.07107      10.6301      8.06226
      6.08276      9.21954      7.07107
      5.09902      7.81025      6.08276
      4.12311      6.40312      5.09902
      3.16228      5.00000      4.12311
      2.23607      3.60555      3.16228
      1.41421      2.23607      2.23607
      1.00000      1.00000      1.41421
      1.41421      1.00000      1.00000
      2.23607      2.23607      1.41421
      3.16228      3.60555      2.23607
      4.12311      5.00000      3.16228
      5.09902      6.40312      4.12311
      6.08276      7.81025      5.09902
      7.07107      9.21954      6.08276
      8.06226      10.6301      7.07107
      9.05539      12.0416      8.06226
      10.0499      13.4536      9.05539
      .....         .....         ....
\end{verbatim}
Let us now ask the question, ``What is the shortest program capable of generating these numbers?''.
These numbers have been generated by the following scheme. First, the 3 particles are put in 3-space, like we usually do:
\bea
\vec{P}_1&=&(1,1,1)+\tau (1,1,1)\\
\vec{P}_2&=& (0,1,0) + \tau (0,1,1) \\
\vec{P}_3&=& (1,1,0) + \tau (1,0,1).
\eea
Then, the distances between the particles are generated,
\bea
r_{12} &=& \sqrt{  (\vec{P}_1-\vec{P}_2 ) \cdot (\vec{P}_1-\vec{P}_2 ) }\\
r_{23} &=& \sqrt{  (\vec{P}_2-\vec{P}_3 ) \cdot (\vec{P}_2-\vec{P}_3 ) }\\
r_{13} &=& \sqrt{  (\vec{P}_1-\vec{P}_3 ) \cdot (\vec{P}_1-\vec{P}_3 ) }
\eea
The parameter labeling ``time'', $\tau$, is just that, a parameter. 

Could we write a program able to infer these equations (in the form of an algorithm, or in a neural network) from the data? This is what our collective and historical brain has been able to do.

\begin{figure}[t!]
\hspace{1.5cm}
\includegraphics[width=10cm]{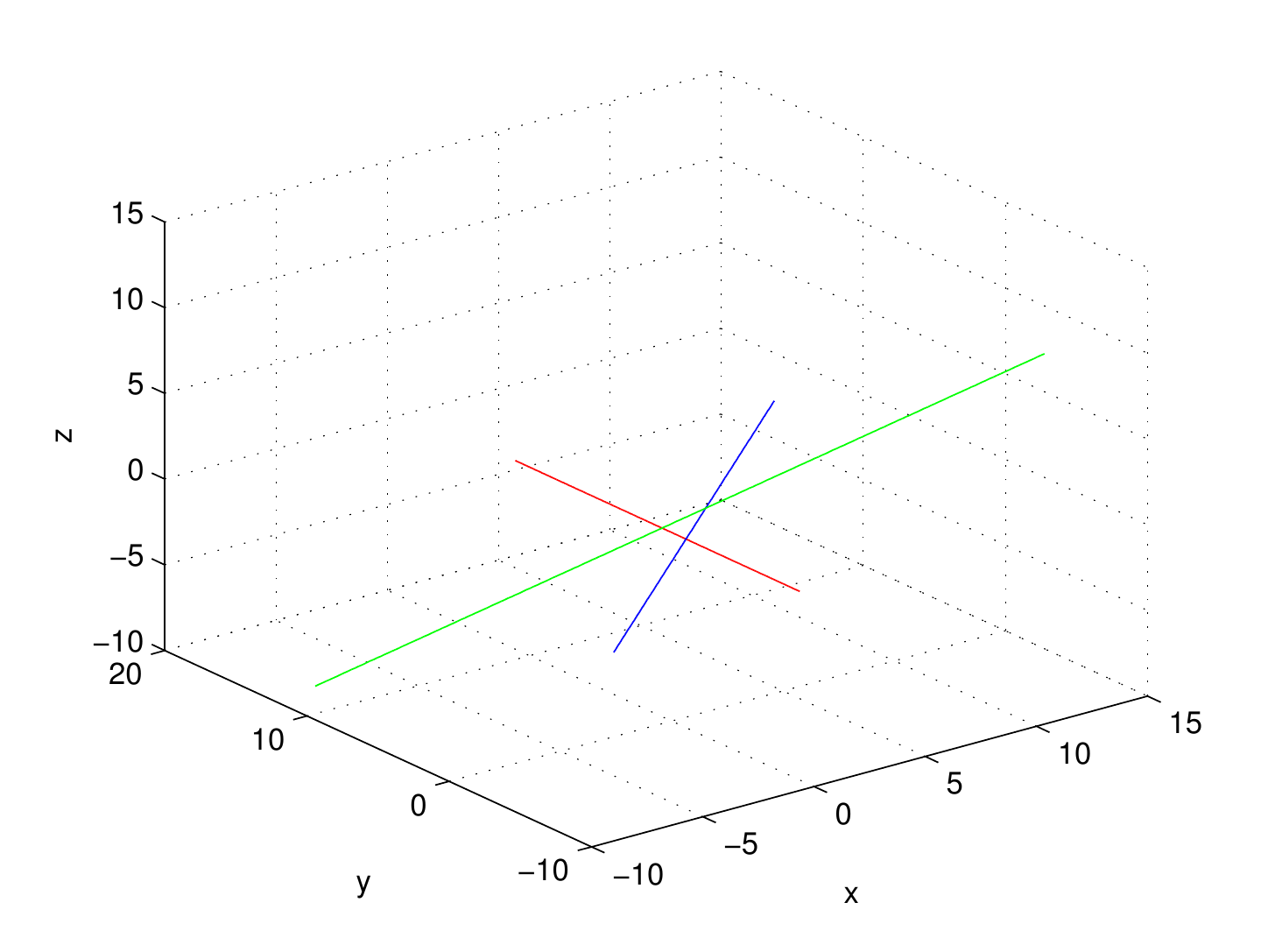}  
\caption{ \label{3particles} Three free particles make a simple universe.}
\end{figure}

\begin{figure}[t!]
\hspace{1.5cm}
\includegraphics[width=10cm]{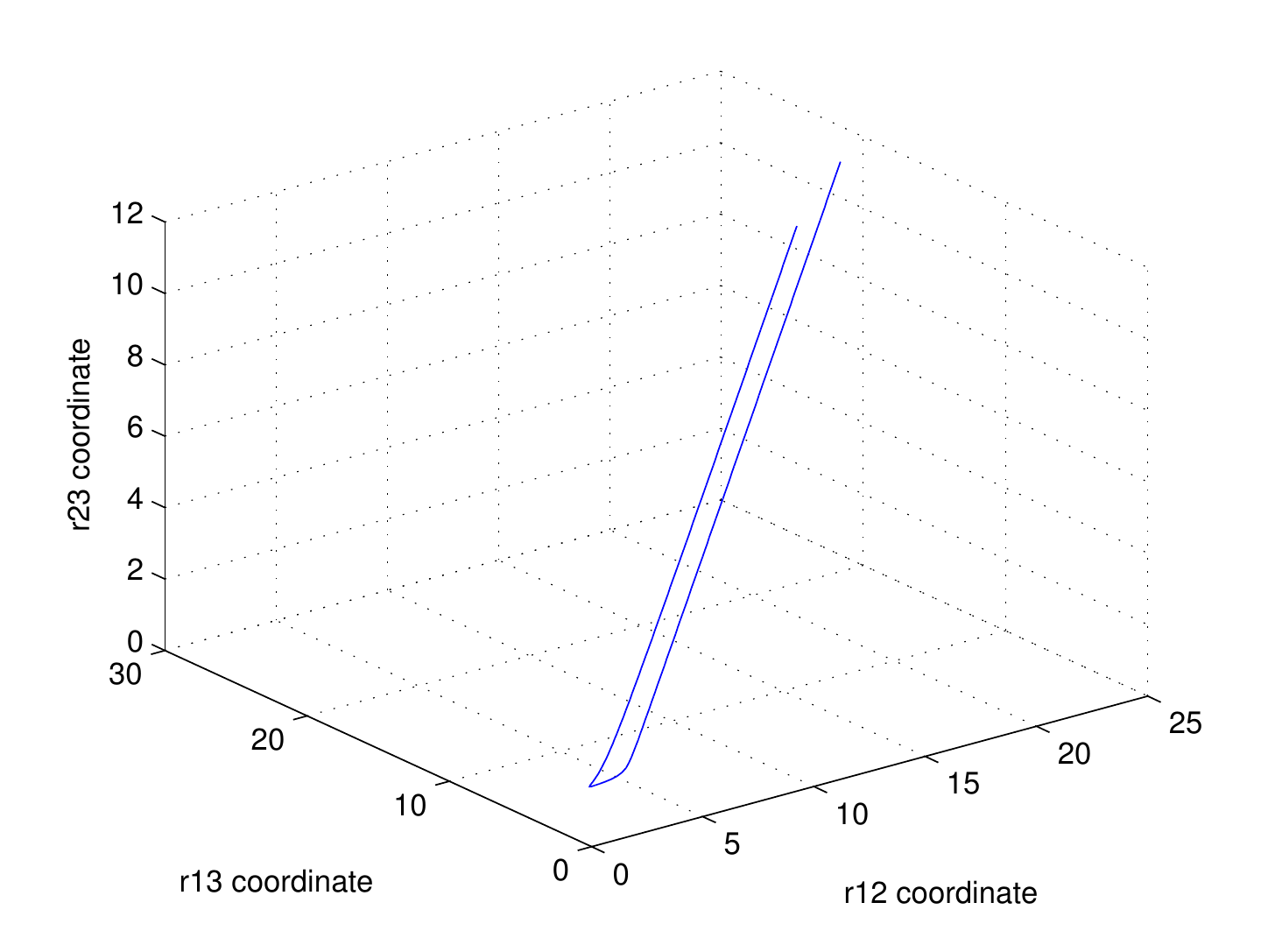}
\caption{ \label{3particlesplat} The Platonia view of the 3 particle universe': only the separation between the particles is relevant. Time arises as a Kolmogorov minimizing construct to simply describe the data.}
\end{figure}


\subsection{Time capsules}

Time capsules, as described by Julian Barbour, are special states in Platonia in which a ``coherent history'' is described. They are in a sense very structured and ``consistent''. Barbour uses the example of geological records which provide consistent accounts of ``historical developments''. I would like to use the term of mutual-information to refer to consistency: consistency can be quantified using the mutual information concept. \medskip

If you take a data stream and it can be highly compressed (i.e., it is simple), they it must follow that subsets of the data have a high mutual information content. \medskip

In another way we can see that the concept of Time is redundant. Time is needed to have the concept of ``Computation''. Computation is intrisically a dynamic process. But as it should be clear that computation does not contribute anything to the information of the initial state: everything is already in the program plus data, which is available at time zero (for those familiar with the concept, a canonical transformation exists which makes the Hamiltonian zero). Both Time and Computation (hence computers) are probably just auxiliary concepts (i.e., gauge concepts in the language of gauge theory).

\subsection{Kolmogorov Complexity and the Wheeler-DeWitt equation}
Can we show that the states selected by the Wheeler-DeWitt equation highlight configurations with a low Kolmogorov Complexity?  What I would like to suggest is that Time Capsules represents minima in the Kolmogorov Complexity of the universe configuration. From this point of view, the quantum Wheeler DeWitt equation would be interpreted as an information-theoretic equation: it selects states which minimize the Kolmogorov Complexity of the universe. \medskip

For example, consider the fact that in the Time Capsule you find yourself now in, mechanical conservation laws are an important holding rule (if this is not the case, you are in a different type of universe!). In the records you will find around you things like mass, the amount of water in a glass, energy and such, are all (at least approximately) conserved. The fact that this happens seems to imply our brains like to exist (or can only exist) in intrinsically simple universes.

\subsection{The arrow of time: from simple to complex}
Why do we remember the past, yet have a very limited picture of the future?  This is a deep and difficult question. I would like to argue that we do have a certain memory of the future (we can predict the next 5 seconds reasonably well, at least in our immediate environment). We can predict that mass will be conserved, that the outside temperature will not change too much..that we will take a plane in the afternoon.... If we follow Barbour's thesis, in which the past is just a structure in the brain (now!), then why do we seem to do a better job in one temporal direction than in the other? Is this a problem related to the Kolmogorov Complexity of the Past and of the Future? It is certainly not a property of the data itself. It is fascinating to think of this problem when only Nows exist! 
\medskip

Another point: why do we only have measurements of the past? Note that the rest is pure data extrapolation (IC, BVP, Newtonian mechanics, Analytic Continuation...). Is it really true? Do we not have imperfect models of the future (extrapolation all the same)?

\subsection{What is computation in Barbour's Platonia?}
The universe is information. In the platonic terms of Barbour, I think it suffices to state that the universe is just a number: the number specifies the state of the system. There is no time, nor space. Only information. 
\medskip

This picture is devoid of space and time, matter, energy.... These are derived emerging concepts. There are patterns in the number, and these can be compressed through the use of these concepts. 
\medskip

It would follow that these patterns are compressed by subportions of the universe...our brains!

\subsection{Data, program, computer}	
One of the concepts that do not sustain a thorough study is the dichotomy of computation, in its traditional terms. A computer is an entity different from the environment: there is the input, the computer, and the output. Such an approach is not useful if one wishes to think of the Universe as a computer. It is possible to define portions of the Universe as computers, but, as usual in cosmology, the extension of the concept to the whole Universe is really troublesome. This situation happens in Quantum Gravity, for instance. When dealing with the whole, a Machian approach is needed. Consider a close system filled with bouncing balls (the billiard universe). The system is in some way computing its own future constantly. But where is the Turing machine, where is the tape? Note that this ``computer'' can be described in a timeless way by the constants of the motion coordinates (or, more directly, by the Hamiltonian and the initial conditions). A subset of billiard balls can be selected as the computer, the rest as the tape, for instance. Both the timeless and the usual time description are in this picture.

\subsection{Working with partial information: Statistical and Quantum Mechanics}

Human brains have access to limited information about the universe. Our sensor systems tap only to data combinations  using a limited number of inputs, etc. This is natural and efficient---why perceive everything? The implications of this simple fact are rather interesting.
\medskip 

Consider Statistical Mechanics (SM). This is a beautiful subset of physics which deals with what knowledge we can extract from physical  so-called {\bf macrostates}. Now, strictly speaking, a  macrostate is not a state at all! When we say that a system is in a macrostate we really mean that we have some limited information about  the microstate the system is in, and that this limited information defines a region of phase space in which the microstate lies. A macrostate is really a set of microstates. \medskip

In fact, the first clue to the crucial role of information in physics, in my opinion, arises from Statistical Mechanics (SM for short). The key concept in SM is microstate and counting. Counting is made possible, especially in the quantum version of SM, because the number of states is countable. Recall that a set is countable if it can be put into a 1-1 relationship with the integers. Being able to count enables the quatification of the number of states in a macrostate. 

\medskip 

Another  important clue, that in my view detaches physics from classical theories of reality comes from quantum mechanics and the concept of the wavefunction. Without going into much detail, I just want to emphasize that in quantum mechanics we are forced to talk about measurements, and reality is described by rules to predict the outcome of experiments. One can and probably should work on quantum mechanics without paying attention to the existence of something "real". The "essence" of reality is not needed to do quantum mechanics. Special Relativity began in a similar way, by questioning things we take for granted in "reality". Concepts such as space-time or simultaneity, which we take for granted, were questioned by Einstein with great profit. He was challenging, in a way, what we take to be the "essence" of reality.

\medskip

To the question of what is information, and its relation to SM, I propose the following answer. Information can be thought of as a set of rules to describe, fully or partially, a microstate. In mathematical terms, information is associated to a set of constraints. In algorithmic or logic terms, these are described by a set of rules. A microstate is fully specified by a complete set of rules. A macrostate is specified by a partial set of rules. The Entropy of a macrostate is the number of bits to specify the missing set of rules to fix the macrostate into a microstate. This is an algorithmic definition.

\medskip
In this way we can talk about the Entropy of a microstate. Let $L$ be the length of the program required to  fully specify a microstate S. The entropy of a macrostate is $E=L-M$, where $M$ is the number of bits used to describe the macrostate. Here we assume that we are working at the KC limit. In normal statistical mechanics the "programming language" is simple. It is a list specifying the position and momenta of each particle, or otherwise specifying the state. In QM things become more complicated due to the fact that all particles are undistinguishable. In such scenarios, the lenght of the list is equivalent to the more familiar concept in which the entropy of a macrostate is defined by the "number of mcro-states" it is really referring to. 
\medskip

The Entropy of a macrostate is the amount of missing information needed to fix it into a micro-state.  

\medskip
According a Noether's theorem, invariances in the action lead to conservation laws. For example, time invariance of the action leads to conservation of energy, and space displacement invariance leads to conservation of momentum. Following the approach explained here, we could say that time and space are concepts derived from the need of human brains to seek invariances. 

\medskip

We can talk here about a more information oriented definition of a self-entity: as self-entity is a semi-isolated information system using a controllable information membrane. It is the ability to control information contamination from the exterior that really makes the self-entity. 

\medskip

As self-entities we  focus (and this is probably not a coincidence) on variables such as temperature, which are rather immune to this lack of knowledge: their probability distribution is very peaked. Well, we deal with averages of things, and the law of large numbers comes to rescue. We can more easily predict. 
\medskip

However, note that when we say that the temperature of the system is T (canonical ensemble), or that the total energy of the system is E (microcanonical ensemble) we are really defining a constraint for the state of the system---a constraint in a huge phase space. Similarly, the entropy of a macrostate is a measure of our lack of information, about the size of the phase space in which the microstate can be given the information we have. If a macrostate is really a set of microstates, entropy is a measure of this set: it is just a count of  how many microstates are in the set. \medskip

Wouldn't it be interesting to work with Algorithmic Complexity instead of Entropy? That is, instead of counting how many microstates there are, we should provide the length of the shortest program providing a description of the microstates.  \medskip

In Quantum Mechanics (QM), it is now being forcibly argued that the state is not a physical state at all: it is a representation of the the information our brain has about the system.  This is very similar to what was just discussed on statistical mechanics. Furthermore,  Quantum mechanics carefully describes what our sensors can and cannot tell us about the universe---think of the Heisenberg principle. As described  in Section~\ref{sensors}, {\em Quantum mechanics is first and foremost a theory describing  what information an observer can access about the universe}.
\medskip

So here is an interesting common aspect to Quantum and classical Statiscal Mechanics. In both Statistical and Quantum mechanics we work with something called a state which is not the state at all (the classical microstate): call it the quantum state or the classical macrostate. In both cases it appears that this state represents our knowledge of the ``real'' state. \medskip

As little detour,  I would like to go back for a moment to Cellular Automata (CAs). As described above in Section~\ref{CA}, there is one rule for elementary binary CAs which is a universal computer and therefore irreducible---rule 110 discussed above. This means that in order to model what will happen in the future, there is no other approach than brute force computation of every single step. Such data streams are therefor incompressible, and their kolmogorov complexity is basically the length of the data stream. In \cite{israeli2004} it is shown that it is possible to ``coarse grain'' such CAs and make them compressible. This means to work with averaged out effective forms of the CA. Something like what we do everyday, when we work with incomplete information.  In other words: even though the original data stream is incompressible (irreducible), if we accept to work with some effective data, we can compress it. I think there is a clear connection here to the discussion above, in which we have highlighted the coarse graining aspects of two fundamental theories. Working with macrostates may allow for the development of models. This may be sufficient for, e.g., survival (unless the essence is in the details).\medskip

Another interesting point is that in Statistical mechanics you have as the most primitive concept the microcanonical ensemble. The microcanonical ensemble is the set of microstates defined by fixing an overall total system energy. It is a rather natural concept, because with time translation invariance in physics laws you have conservation of energy. We could think of it as the statistical mechanics framework of the universe. However, if you are interested about the physics of a subsystem which by itself may not conserve energy (because of interaction with the rest of the system), you work with the canonical ensemble: this is done by averaging out over all states outside the subsystem of interest (``tracing out''). Out of this very interesting process, the canonical ensemble is born and the canonical partition function is defined. This looks a lot like a the Hamiltonian propagator in QM, except that time is changed by a complex quantity, $1/k\beta$ (see the lucid discussion in {\em Feynman and Hibbs} \cite{FeynmanHibbs}). The connection between statistical mechanics and classical dynamics (the Hamiltonian) has to do, no doubt, with the ergodic hypothesis. In SM, the Hamiltonian is our ``chaperon'' or escort of phase space: it takes us, eventually, everywhere where the energy in constant. Thus, we can do time averages for constant energy phase space averages---which is what we want in the microcanonical ensemble.  All accessible microstates are equally probable over long period of time, or, equivalently,  time averaging and averaging over the statistical ensemble are the same. \medskip

In the SM of a free gas, for instance, the Hamiltonian is the sum of the Hamiltonians for each particle.  \medskip

A related fact: Planck's constant has the units of action. Action units are volume units in phase space. Both in QM and SM, to compute the Feynman path integral or the statistical partition function, the key operation is a sum over volumes in a large phase space. In SM the space is large because you have many particles. \medskip

In the QMs of a single free particle you sum over all possible paths in time. A phase space path integral is really a sum over a super-phase space though (think about the ergodic hypothesis!). In the QM phase space path integral, a single particle moving in time emulates, mathematically speaking, a statistical ensemble of free particles. 
The resulting expression for the propagator is symbolized 
by
\begin{equation}
U =  \int  DxDp \; e ^{i \int_{t_i}^{t_f}  pdx - Hdt} =  \int  DxDp e^{ i \int_{t_i}^{t_f}
dt(p{\dot{x} }- H)},
\end{equation}
where the measure here means
\begin{equation}
DxDp \equiv { dp_0  \over 2 \pi}  \prod_{i=1}^N {dx_i dp_i \over 2 \pi}.
\end{equation}
The QM measure for the integral is that of a phase space in which there are $N$ particles. The propagator is a weighted sum over this (super) phase space.  \medskip

No wonder there is a strong (mathematical) connection between the QM and SM---as also discussed in {\em Feynman and Hibbs} \cite{FeynmanHibbs}. An important difference is the imaginary term in QM. This means, somehow, that instead of having on probability you have two numbers. And the, of course, there are all the aspects of interpretation of the amplitude. \medskip 
 
Time is in this picture from the beginning, however, because in the microcanonical ensemble you define a total energy of the universe, and energy and time are strongly related (canonical conjugates). \medskip

By the way, this ``canonical'' process of tracing degrees of freedom out is also in discussions by Barbour on how inertia arises from the distribution of matter in the universe. The total universe is postulated to have fixed (zero) angular momentum, but local subsystems do not need to conserve momentum. Inertia can be seen to arise from this ``effective'' theory. 

\clearpage

\section{CLOSING}
The underlying theme of this set of notes on the theme of Kolmogorov Complexity has been  the relationship between Brain and  Reality, or Neuroscience, Information theory and physics. I have argued that we cannot really progress in our understanding of Reality without understanding the Reality Machine, the Brain. I have also argued that compression is the main task of the Brain, and that from compression the picture of Reality arises.
\medskip

The discussed  conceptual framework can be summarized as follows:
\begin{itemize}
\item The brain receives all its information flows from the senses, and constructs models for the incoming data. All we can assert for certain is that Information exists. The brain is really a Turing machine, with the tape head role being played by active sensors and actuators.
\item The universe is a dynamical system providing the infrastructure for the evolutive emergence of Turing machines.
\item These models  generate the concepts we associate with reality. They are reality as far as the brain is concerned. 
\item It is a key aspect that we have access only to a limited set of information, that we cannot really identify the Universe microstate. This is due to our limited sensing powers, and this in turn is driven by evolution.  This limitation is no doubt closely related to the information regulation rules in  Quantum Mechanics and Statistical Mechanics which we have ``discovered'' and their focus on working with partial information. As organisms, we have developed sensors and algorithms focusing on limited information availability, e.g., targeting quantities which are stable (``the temperature of a system'') and useful to us (``the position of the particle''). 
\item The driving force in the brain and in model building is compression. Thereby arise concepts such as space and time, ``the Great Simplifier''. If a model and its concepts can compress the data, then we assign a level of reality to them. Useful concepts and models are real. 
\item Evolution and Natural selection play an important role. Models for the universe evolve. Our brains themselves have evolved as compressors. Compression is power. DNA itself contains compressed representations of ``reality''.
\item In a more speculative and platonic vein, the universe can be conceptualized as a number. This number can be thought of as representing the dynamical data stream.  In this number (perhaps $\pi$!) there are patterns, and these interact. Some of these are Turing machines.  Somehow, some portions of the Universe contain information on the rest. Our brains do that, DNA does that. As argued by Barbour, this is the mechanism through which the concept of time arises. The Nows contain information about the Pasts.
\end{itemize}

Here I will try to list a series of more concrete questions to address as part of this research program:
\begin{enumerate}
\item How can we in practice differentiate and  quantify complexity? Can we differentiate the DNA sequence from a live organism from a random sequence?
\item How much can we compress the DNA sequence for ``Drosophila''? What is its Kolmogorov Complexity? Does this correlate in some sense with the complexiy of its ``effective universe of interaction''?
\item Can we discover the encoding mechanisms of the brain by studying learning processes as described above? Does the brain encode in a specific way, optimized for our environment? How does the Pattern Lock Loop machine work? Can we replicate it in a simulator or electronic hardware?
\item What experiments can we design to test the role of KC and ``pattern resonance'' in brain function?
\item Is there any fundamental  difference between Reality and Virtual Reality? The area of Presence is precisly addressing this question by defining experimental benchmarks of what type of information baths feel real.
\item What is the Kolmogorov Complexity of the Universe (this is a difficult one, but physicists are working on it)?
\item What is the Kolmogorov Complexity of a dream universe?
\item Can we measure the coherence (which can also be defined in terms of the Kolmogorov Complexity of the entire set of programs in an organism) of an organism and correlate it with age?
\item Can we show that the states selected by the Wheeler-DeWitt equation highlight configurations with a low Kolmogorov Complexity? In a sense they do, because they satisfy that equation. What makes that equation special in the complexity context?
\item Can we devise pattern lock loop programs? A possible approach may be provided by CAs (rule 110) and GAs. This is an interesting, practical area of work with ``real'' potential applications.
\end{enumerate}

\section*{Acknowledgments}
I wish to thank Ed Rietman, Walter van de Velde, Carles Grau, Josep Marco, Julian Barbour, Bert Kappen, Isabel Bouhom and Rodrigo Quiroga for useful comments during the writing of this essay. This research has been funded by Starlab.

\end{document}